\documentclass[prd,preprint,showpacs,preprintnumbers,amsmath,amssymb]{revtex4}

\usepackage{graphicx}% Include figure files
\usepackage{dcolumn}% Align table columns on decimal point
\usepackage{bm}% bold math

\newcommand{\nl}{\nonumber \\ }

\newcommand{\order}{{\cal O}}

\newcommand{\be}{\begin{equation}}  
\newcommand{\ee}{\end{equation}}  
\newcommand{\bear}{\begin{eqnarray}}  
\newcommand{\eear}{\end{eqnarray}}  
\newcommand{\ba}{\begin{array}}  
\newcommand{\ea}{\end{array}}

\newskip\humongous \humongous=0pt plus 1000pt minus 1000pt

\newif\ifdtup

\def\oldreffmt#1{\rlap{[#1]} \hbox to 2\parindent{}}

\def\figfmt#1{\rlap{Figure {#1}} \hbox to 1in{}}  
  
\def\slash#1{#1\!\!\!/\!\,\,}  
\def\beq{\begin{equation}}  
\def\eeq{\end{equation}}  
\def\bea{\begin{eqnarray}}  
\def\eea{\end{eqnarray}}

\def\bq{\begin{quote}}  
\def\eq{\end{quote}}

\relax  

\newdimen\tdim  
\tdim=\unitlength  
\def\bar{\overline}

\begin{document}  

\preprint{EFI Preprint 09-05}
\title{
Low energy analysis of $\nu N \to \nu N \gamma$ in 
the Standard Model 
}

\author{
Richard J. Hill
}

\email{
richardhill@uchicago.edu
}

\affiliation{\vspace{0.2in}
Enrico Fermi Institute and Department of Physics \\
The University of Chicago, Chicago, Illinois, 60637, USA
}

\date{\today}

\begin{abstract} 
The production of single photons in low energy ($\sim$ 1~GeV) 
neutrino scattering off nucleons
is analyzed in the Standard Model.  
At very low energies, $E_\nu \ll {\rm GeV}$, a simple description of the 
chiral lagrangian involving baryons and arbitrary 
$SU(2)_L\times U(1)_Y$ gauge fields is developed. 
Extrapolation of the process into the $\sim 1-2 \, {\rm GeV}$ 
region is treated in a simple phenomenological model.  
Coherent enhancements in compound nuclei are studied.  
The relevance of single photon events as a background to experimental 
searches for $\nu_\mu \to \nu_e$ is discussed.  In particular, 
single photons are a plausible explanation for excess events observed 
by the MiniBooNE experiment. 
\end{abstract}

\maketitle

\section{Introduction}

Recently, it was shown that a careful gauging of the low energy
chiral lagrangian of QCD leads to new anomalous interactions,
``pseudo Chern Simons'' (pCS) interactions, 
between the $Z$-boson, the photon and strongly coupled vector mesons, 
such as the spin-$1$ $\omega$ meson, taking the form
$\propto \epsilon^{\mu\nu\rho\sigma}\omega_\mu Z_\nu F_{\rho\sigma}$~\cite{HHH1}.
At low energies, this interaction implies anomalous
processes involving neutrinos and $\gamma$'s 
in the presence of nuclear fields, e.g. contributing to 
$\nu N \to \nu N \gamma$.   

Previous simple estimates suggest that
these effects may be at work in experimental
configurations such as MiniBooNE~\cite{MiniBooNE}.
They can lead to a peculiar enhancement in the
appearance of ``electrons'' in a $\nu_\mu$ beam, where
the hard $\gamma$'s are actually faking electron 
Cerenkov signatures \cite{HHH2}.
This phenomenon may also play a role in
astrophysical applications such as in neutron star cooling
and supernova dynamics.
The present paper widens the analysis of the phenomenology
of such novel interactions, focusing on laboratory detection.

A rigorous discussion of photon production by the weak neutral
current can be obtained at low neutrino energies using a chiral
lagrangian description.   
Section~\ref{sec:chiral} reviews the key aspects of the 
chiral lagrangian in the presence of baryons, and
extends the usual formalism to describe general vector and axial-vector
couplings of neutral electroweak fields.  
Terms induced by the pCS interactions appear at 3-derivative order, 
and are described by 
an essentially unique new operator appearing at this order.   
The effects of this interaction may be accessible in processes such as 
radiative neutrino scattering on baryons, or in certain parity violating
observables%
\footnote{
For example, in analogy to the discussion in \cite{HHH2}, 
the pCS interactions induce an anapole moment at finite baryon density. 
}. 

Section~\ref{sec:model} investigates in more detail 
the processes of Compton-like scattering (with a weak vector or 
axial-vector current replacing one of the photons), 
$t$-channel $\omega(780)$ resonance
exchange, and $s$-channel $\Delta(1232)$ resonance production.   
These processes serve to fix the normalization of the relevant interactions 
appearing in the chiral lagrangian, and provide a form-factor model 
to extrapolate the phenomenological predictions into the 
$E_\nu \sim 1\, {\rm GeV}$ energy range. 
An explicit computation of the competing contributions involving
the axial-vector weak current reveals 
the significance of the operator induced by the above-mentioned pCS term: 
it is the unique interaction appearing through 3-derivative order 
in the chiral lagrangian that leads to a coherent coupling of one axial-vector gauge field 
and one vector gauge field to baryons. 
A significant contribution to the coefficient of this operator
is also induced from the $\Delta$ resonance.
Pertinent details of the $\Delta$ coupling to baryons and gauge 
fields are reviewed.  The relevance 
of ``off-shell'' parameters for the $\Delta$ couplings to nucleons and
vector currents is investigated, and simplifications in the 
formal large-$N_c$ limit are discussed.     

Section~\ref{sec:pheno} presents cross section estimates for 
single-photon production in neutrino-nucleon 
scattering, i.e., 
$\nu(\bar{\nu}) n \to \nu(\bar{\nu}) n \gamma$, 
$\nu(\bar{\nu}) p \to \nu(\bar{\nu}) p \gamma$. 
Section~\ref{sec:coh} discusses several aspects of coherent scattering 
on compound nuclei, and derives scaling laws for the coherent 
cross sections in the limit of a large nucleus.  
Section~\ref{sec:conclusion} concludes with a brief discussion on
the relevance of single photon production as a background to 
experiments searching for $\nu_\mu \to \nu_e$ oscillation, and an 
outline for future work.  

\section{Chiral lagrangian for nonrelativistic baryons and
electroweak gauge fields \label{sec:chiral}} 

Consider physical processes involving a single nucleon 
interacting with electroweak fields at energies small compared to 
typical hadronic scales $\sim 1\, {\rm GeV}$.   
The relevant dynamical fields include the 
pions, identified as Nambu-Goldstone bosons (NGBs) of spontaneously broken 
chiral symmetry; the nucleons; and the electroweak fields.  To begin, 
let us concentrate on processes involving almost-stationary nucleons, 
so that a nonrelativistic 
expansion is appropriate.  The task is to build the most general 
effective lagrangian from these fields, order by order in the 
small parameters $E/ m_N \sim E/ 4\pi f_\pi \ll 1$.
The full nonrelativistic expansion is cumbersome~\cite{Jenkins:1990jv}, 
and for a tree-level analysis it is simpler to work in a 
manifestly Lorentz-invariant form of the Lagrangian. 
The basic formalism is common to 
previous analyses~\cite{Gasser:1987rb,Weinberg:1991um,Bernard:1995dp}, 
which however neglect a systematic treatment of the $U(1)$ factors.
Care will be taken  to include the
full $SU(2)_L \times SU(2)_R \times U(1)_V$ symmetry that 
encompasses the Standard Model gauge group.   

\subsection{Fields in the effective theory}
Let us first make a few comments on the treatment of NGBs, 
nucleons and electroweak gauge fields in the effective theory. 
This will also serve to introduce notations and conventions. 
As usual, the pions are collected into the $SU(2)$ matrix field
(considering for simplicity just $n_f=2$ light flavors), 
\be
U = \exp[ (2i/f_\pi) \pi^a T^a ] \,, 
\ee
with $T^a = \tau^a/2$ and $f_\pi \approx 93\,{\rm MeV}$.  

The nucleons $p$ and $n$ compose a field 
\be
N = \left( 
\begin{array}{c} 
p \\ n \end{array} \right) \,,
\ee
that transforms linearly under isospin.
Extending linear $SU(2)_V$ isospin 
transformations to general $SU(2)_L \times SU(2)_R$
requires a nonlinear transformation law that reduces to the
linear one when restricted to the unbroken $SU(2)_V$ subgroup~\cite{CCWZ}.
This is achieved by introducing a field $\xi$ such that
\be
\xi^2 = U \,.
\ee
Consider an $SU(2)_L\times SU(2)_R \times U(1)_V$ transformation of the underlying quarks:  
$\psi_{L,R} \to e^{i\epsilon_{L,R}}\psi_{L,R}$.  
Then $U$ is defined to transform as:
\be
U \to e^{i\epsilon_L} U e^{-i\epsilon_R} \,,
\ee
and we can further define a quantity $\epsilon'$ 
by requiring the corresponding transformation
law of $\xi$: 
\be
\xi \to e^{i\epsilon_L}\xi e^{-i\epsilon'} = e^{i\epsilon'} \xi e^{-i\epsilon_R} \,. 
\ee
Finally, we define $N$ to transform as
\be
\label{eq:nonlin}
N \to \exp\left[ 
i\left( \hat{\epsilon}'+\frac32{\rm Tr}(\epsilon') \right) 
\right] N \,.  
\ee
Here
$\hat{M} \equiv M - \frac12 {\rm Tr}(M)$ 
denotes 
the traceless part of a general $2\times 2$ matrix $M$. 
Note that the factor of $3$ in front of ${\rm Tr}(\epsilon')$
reflects the underlying quark content of the nucleons.

Now consider the usual $SU(2)_L\times U(1)_Y$ gauge subgroup of the Standard Model. 
At the physical value of the electroweak coupling constant $g_2$,
and Higgs vacuum expectation value $v_{\rm weak}$,  
the $W$ and $Z$ bosons are very heavy, $m_{W,Z} \sim g_2 v_{\rm weak}$, 
and are integrated out of the theory. In order to keep track of
gauge-invariance constraints that survive at low-energy, 
first consider the limit $v_{\rm weak} \gg f_\pi$ fixed and 
$g_2$ small.  In this case, $W$ and $Z$ still eat mostly the 
Higgs NGBs (not the pions) but have small mass.  At the end of 
the analysis, $g_2$ can then be set to its physical value
and the massive vector bosons integrated out to induce operators 
involving only light fields (such as leptons).  

\subsection{Covariant building blocks and effective lagrangian}
 
Let $A_L^\mu$ and $A_R^\mu$ represent weakly coupled gauge fields (or external fields) 
acting on left- and right-handed quark fields,
and let us define 
\begin{align}
\tilde{A}_L^\mu &\equiv \xi^\dagger ( i\partial^\mu + A_L^\mu ) \xi \,, \nl
\tilde{A}_R^\mu &\equiv \xi ( i\partial^\mu + A_R^\mu ) \xi^\dagger \,.
\end{align}
These objects have the transformation laws, 
\begin{align} 
\tilde{A}_L^\mu &\to e^{i\epsilon'} (\tilde{A}_L^\mu + i\partial^\mu)e^{-i\epsilon'} \,, \nl 
\tilde{A}_R^\mu &\to e^{i\epsilon'} (\tilde{A}_R^\mu + i\partial^\mu)e^{-i\epsilon'} \,. 
\end{align} 
Under parity, the fields transform as 
\be
\xi \leftrightarrow \xi^\dagger \,, \quad 
A_L \leftrightarrow A_R \,,
\ee
so that 
\be 
\tilde{A}_L \leftrightarrow \tilde{A}_R \,. 
\ee
Fields with definite parity are the linear combinations 
\begin{align}
\tilde{V} &\equiv \frac12 \left( \tilde{A}_L + \tilde{A}_R \right) \,, \nl
\tilde{A} &\equiv \frac12 \left( \tilde{A}_L - \tilde{A}_R \right) \,,
\end{align} 
transforming under parity as vector and axial-vector, respectively. 
Under a gauge transformation we have
\begin{align} 
\tilde{V}_\mu 
&\to e^{i\epsilon'}( \tilde{V}_\mu + i\partial_\mu ) e^{-i\epsilon'} \,, \nl
\tilde{A}_\mu 
&\to e^{i\epsilon'} \tilde{A}_\mu e^{-i\epsilon'} \,. 
\end{align} 

Eq.~(\ref{eq:nonlin}), shows that 
in order to build invariant operators involving the 
nucleon field, the trace component of the vector field should appear as: 
\begin{align}\label{eq:Vnonlin}
\tilde{V}'_\mu \equiv \hat{\tilde{V}}_\mu + \frac32{\rm Tr}(\tilde{V}_\mu) 
&\to 
 \exp\left[ 
i\left( \hat{\epsilon}'+\frac32{\rm Tr}(\epsilon') \right) 
\right]
\big( 
\tilde{V}'_\mu + i \partial_\mu
\big)
\,
 \exp\left[ 
-i\left( \hat{\epsilon}'+\frac32{\rm Tr}(\epsilon') \right) 
\right] \,.
\end{align}
Thus $\tilde{A}$ is a covariantly transforming axial-vector field, 
and $\tilde{V}'$ is a vector field that can be used to form the covariant 
derivative acting on nucleon fields,  
\be
i\tilde{D}_\mu = i\partial_\mu + \tilde{V}'_\mu \,. 
\ee
Explicitly, the leading expansions for vector and axial-vector fields are
\begin{align}
\tilde{V}'_\mu &= {g_2 \over 4} 
\left( \begin{array}{cc} 
{1\over c_W}(1-4s_W^2) Z_\mu & \sqrt{2} W^+_\mu \\
\sqrt{2} W^-_\mu & -{1\over c_W}Z_\mu 
\end{array} \right)
+  
e\left( \begin{array}{cc} 
A^{\rm e.m.}_\mu & 0 \\
0 & 0 
\end{array} \right) + \dots \,,
\nl 
\tilde{A}_\mu &= 
{g_2 \over 4} 
\left( \begin{array}{cc} 
{1\over c_W}Z_\mu & \sqrt{2} W^+_\mu \\
\sqrt{2} W^-_\mu & -{1\over c_W}Z_\mu 
\end{array} \right)
-{1\over 2 f_\pi} 
\left( \begin{array}{cc} 
\partial_\mu \pi^0  & \sqrt{2} \partial_\mu \pi^+ \\
\sqrt{2} \partial_\mu \pi^- & -\partial_\mu \pi^0
\end{array} \right)
+ \dots \,,
\end{align}
where dots denotes terms with two or more pions. 
The notation $s_W=\sin\theta_W$, $c_W=\cos\theta_W$ is
used throughout, where $s_W^2=0.231$. 

It is now straightforward to write down the
Lagrangian working order by order in derivatives. 
Consider the expansion through three-derivative order, 
where the interesting effects of pCS terms make their appearance. 
In the one-baryon sector, 
\be\label{rexpand}
{\cal L} = m_N {\cal L}^{(0)} + {\cal L}^{(1)} 
+ {1\over m_N}{\cal L}^{(2)} 
+ {1\over m_N^2}{\cal L}^{(3)} + \dots \,, 
\ee
where 
for convenience the mass scale in the expansion 
is defined as the nucleon mass, $m_N\approx 940\,{\rm MeV}$.
For notational simplicity, the tildes on fields 
are dropped in the remainder of this section.   
Using hermiticity and enforcing invariance under parity and 
time-reversal, 
and making use of the leading-order equations of motion 
\begin{align}
(i\slash{D} - m_N) N &\sim 0 \,, \nl
[D_\mu , {A}^\mu ] &\sim 0 \,,
\end{align}
the result is:
\begin{align}\label{eq:expansion}
{\cal L}^{(0)} &= - c^{(0)} \bar{N} N \,, \nl
{\cal L}^{(1)} &= \bar{N} \big[ c^{(1)}_1 i\slash{D} 
- c^{(1)}_2 \slash{A} \gamma_5 
\big] N \,, \nl
{\cal L}^{(2)} &= \bar{N} \big[
- 
c^{(2)}_1 \frac{i}{2} \sigma^{\mu\nu} {\rm Tr}( [iD_\mu \,, iD_\nu ] ) 
- 
c^{(2)}_2 \frac{i}{2} \sigma^{\mu\nu}
\tau^a {\rm Tr}( \tau^a [iD_\mu \,, iD_\nu]  
)
+ \dots 
\big] N 
\,, \nl
{\cal L}^{(3)} &= \bar{N} \big[ 
c^{(3)}_1 \gamma^\nu [ iD_\mu , {\rm Tr}( [iD^\mu , iD_\nu] )]
+ c^{(3)}_2 \gamma^\nu [ iD_\mu , \tau^a {\rm Tr}\big( \tau^a [iD^\mu , iD_\nu] \big)]
\nl 
& \qquad
+ c^{(3)}_3 \gamma^\nu \gamma_5 [iD_\mu ,  [iD^\mu , A_\nu ] ]
\nl
& \qquad
+ c^{(3)}_4 i \epsilon^{\mu\nu\rho\sigma} \gamma_\sigma 
{\rm Tr}\big( \{ A_\mu , [iD_\nu , i D_\rho ] \} \big)
+ c^{(3)}_5 i \epsilon^{\mu\nu\rho\sigma} \gamma_\sigma 
\tau^a {\rm Tr}\big( \tau^a \{ A_\mu , [iD_\nu , iD_\rho] \} \big)
\nl
&\qquad
+ c^{(3)}_6 \gamma^\nu \gamma_5 [ [iD_\mu \,, iD_\nu] \,, A^\mu ]  
+ c^{(3)}_7 
{1\over 4m_N^2}\gamma^\nu \gamma_5 \big\{ 
 [ [iD_\mu \,, iD_\nu] \,, A_\rho ] , \{ iD^\mu\,, iD^\rho\} \big\}
+ \dots \big] N
\,.
\end{align} 
The dots in ${\cal L}^{(2)}$ and ${\cal L}^{(3)}$ 
denote terms containing more than one $A$ field. 
The simplification ${\rm Tr}(A)=0$ has been made,  
which is sufficient for Standard Model applications.  
When restricted to isovector gauge couplings, 
these expressions are equivalent to previous 
results at second~\cite{Gasser:1987rb} 
and third~\cite{Fettes:1998ud} order.  
The treatment of $U(1)$ factors encoded by (\ref{eq:nonlin}) 
and (\ref{eq:Vnonlin})
allows also the isoscalar components of both the 
photon and $Z$ boson to be incorporated systematically%   
\footnote{
The $U(1)_V$ factor has either been ignored~\cite{Gasser:1987rb} or 
added by hand~\cite{Bernard:1995dp} in previous discussions where either only 
pions, or only pions and photons were relevant to the discussion.   
}.
In (\ref{eq:expansion}),
field redefinitions have been used to arrange operators in a manner that 
allows a straightforward interpretation in terms of 
a vector meson dominance model.    
Note that in expanding in derivatives, care must be taken 
to notice that the time component counts as order 
one,
\be
i\partial_0 N \sim \order(m_N) N \,,
\ee
and hence in a relativistic and gauge-invariant description, 
covariant derivatives $D_\mu$ acting on the nucleon field 
cannot be viewed as power suppressed.  
Within the approximations of interest (three derivative order, and
no more than one axial-vector field), the final term in 
(\ref{eq:expansion}) is the only new operator induced by this 
subtlety~\cite{Fettes:1998ud}.

Despite appearances, the expansion (\ref{eq:expansion}) is remarkably 
simple when applied to the problem at hand.
The leading coefficients define the mass and field normalization, 
$c^{(0)} = c^{(1)}_1 = 1$. 
The remaining first and second-order coefficients are 
related at tree level to well-known low-energy observables. 
Coefficient $c^{(1)}_2$ is the axial-vector coupling to nucleons: 
$c^{(1)}_2 \equiv g_A \approx 1.26$. 
Coefficients $c^{(2)}_{1,2}$ represent 
the isoscalar and isovector anomalous magnetic moments 
(in units $1/2 m_N$), 
which at tree level would be:
$c^{(2)}_1 \approx\frac14 ( a_p + a_n )$,
$c^{(2)}_2 \approx \frac14 ( a_p - a_n )$, 
where $a_p=1.79$ and $a_n=-1.91$. 
For the third-order constants, $c^{(3)}_1$ and 
$c^{(3)}_2$ correspond to form-factor corrections to the 
leading vector couplings; in a vector dominance approximation,
$c^{(3)}_1 \approx \frac12 {m_N^2/ m_\omega^2 }$,
$c^{(3)}_2 \approx \frac12 {m_N^2/ m_\rho^2 }$,
and similarly for axial-vector coupling, 
$c^{(3)}_3 \approx -g_A {m_N^2 / m_{a_1}^2 }$.

We are left finally with $c^{(3)}_{4,5,6,7}$.  
The coefficients $c^{(3)}_6$ and $c^{(3)}_7$ 
will not be relevant, since the corresponding operators vanish for neutral gauge fields
such as the $Z^0$ and the photon.  
The coefficients $c^{(3)}_{4,5}$ contain the low-energy manifestation 
of the $\omega Z dA$ and $\rho Z dA$  vertices studied in Ref~\cite{HHH1,HHH2}, 
after integrating out $\omega$ and $\rho$:
\begin{align}
c^{(3)}_4(\omega) &\sim  {9\over 32\pi^2} {g'^2 m_N^2 \over m_\omega^2} \sim 1.5, \nl
c^{(3)}_5(\rho) &\sim {1\over 32\pi^2} {g^2 m_N^2 \over m_\rho^2} \sim 0.2 \,. 
\end{align}
The conventions $\epsilon^{0123}=-1$, 
$g^{\mu\nu}={\rm diag}(1,-1,-1,-1)$ are used throughout.   
The sign of $c_4^{(3)}(\omega)$
is fixed by noticing that the baryon current has 
divergence 
$\partial_\mu J^\mu = -(eg_2/8\pi^2 c_W) \epsilon^{\mu\nu\rho\sigma} \partial_\mu A^{\rm e.m.}_\nu \partial_\rho Z_\sigma + \dots$, 
and enforcing that $\omega$ couples equally to all parts of the 
baryon current%
\footnote{
An errant minus sign appears in Eq.(76) of Ref.~\cite{HHH1}.   This is
a transcription error from the (correct) Eq.(74) in the same reference.  
}.
We will see below that $c^{(3)}_4$ also receives significant contributions 
from $\Delta(1232)$.  

This discussion allows us to make more precise the significance of
the pCS terms on low-energy physics: the operators that the pCS 
terms match onto are the only essentially new terms 
(i.e., besides terms representing form factor corrections to leading operators)
in the baryon chiral lagrangian coupled to neutral vector and axial-vector fields
through three derivative order.   
Such operators have the special property that they can 
act coherently on adjacent nucleons, yet involve axial-vector gauge fields.

\subsection{Nonrelativistic expansion} 

In the relativistic formalism, there is not an explicit 
scale separation, for two reasons.  First, 
the relativistic nucleon spinor contains suppressed terms
\be
{u_s(k)\over \sqrt{2m_N}} \sim 
\left(
\begin{array}{c}
\chi_s \\ 0 
\end{array} 
\right) 
+ {\bm{\sigma}\cdot\bm{k} \over 2m_N } 
\left(
\begin{array}{c}
0 \\ \chi_s
\end{array} 
\right) + \dots \,.
\ee
Second, there is an intermediate scale $Q^2 \sim |\bm{p}| m_N$ that arises in
diagrams such as Fig.~\ref{fig:compton}:
\be
{1\over (p+k)^2 - m_N^2 } 
= {1\over 2p\cdot k + p^2} \sim {1\over 2 |\bm{p}|m_N} \,.
\ee
For the present tree-level analysis, 
it is not technically necessary to make the scale separation 
\be
|\bm{p}|^2 \ll |\bm{p}|m_N \ll m_N^2 
\ee
explicit.  However, since it is instructive to see how the various contributions to
$\nu N \to \nu N \gamma$ arise in such a formalism, an outline 
is presented here. 

\subsubsection{Nonrelativistic fields and effective lagrangian}

For nonrelativistic nucleons, the antiparticle components of the nucleon 
field are integrated out, and we deal with two-component (Pauli) spinor fields
$\chi_p$, $\chi_n$, in place of four-component (Dirac) spinor fields.  
This expansion does not affect the chiral transformation properties, and 
the nonrelativistic isodoublet nucleon field transforms as before:
\be
\hat{N} = \left( 
\begin{array}{c} 
\chi_p \\ \chi_n \end{array} \right) 
\to 
e^{i\epsilon'} \hat{N} \,, 
\ee
where hats are here used to denote nonrelativistic fields. 
(For a proper treatment of the $U(1)$ factors, the modification 
(\ref{eq:nonlin}) is understood.)

Invariant combinations can be formed from the operators: 
\be
D_0\,, A_0 \,,
D_i \,,
A_i \,,
\ee 
sandwiched between $\hat{N}^\dagger( \dots )\hat{N}$.  
Let us denote the expansion as 
\be\label{nrexpand}
\hat{\cal L} = m_N \hat{\cal L}^{(0)} 
+ \hat{\cal L}^{(1)} 
+ {1\over m_N}\hat{\cal L}^{(2)} 
+ {1\over m_N^2}\hat{\cal L}^{(3)} + \dots \,. 
\ee   
Lorentz symmetry is broken by the nonrelativistic limit, but
parity, time-reversal and rotation invariance of the strong interactions 
is preserved. 
In particular, 
the fields $D_i$ and $A_0$ are odd under parity, and so must appear an even 
number of times.  
 
At zeroth order,
\be
\hat{\cal L}^{(0)} = \hat{c}^{(0)}  \hat{N}^\dagger \hat{N} \,. 
\ee
By an appropriate field redefinition ($\hat{N}\to e^{-im_Nt} \hat{N}$), this term can be 
removed, $\hat{c}^{(0)} \equiv 0$.
At first order,
\be\label{lhat1}
\hat{\cal L}^{(1)} = \hat{N}^\dagger(
\hat{c}^{(1)}_1  iD_0 
+ \hat{c}^{(1)}_2 \bm{\sigma}\cdot \bm{A} 
)\hat{N} \,.
\ee
The normalization of the field $\hat{N}$ can be chosen such that 
$\hat{c}^{(1)}_1 \equiv 1$ 
and then $\hat{c}^{(1)}_2$
determines the tree-level axial-vector coupling. 
At second order,
\begin{align}\label{lhat2}
\hat{\cal L}^{(2)} &=  \hat{N}^\dagger\bigg(\!\!
 - \frac12  \hat{c}^{(2)}_1 \bm{D}^2
+ \hat{c}^{(2)}_2 \epsilon^{ijk} \sigma^i {\rm Tr}( D_j D_k ) 
+ \hat{c}^{(2)}_3 \epsilon^{ijk} \sigma^i \tau^a {\rm Tr}( \tau^a D_j D_k ) 
+ i \hat{c}^{(2)}_4 \{\bm{\sigma}\cdot\bm{D}, A_0\} 
\!\!\bigg) \hat{N} \,,
\end{align}
where the leading order equation of motion,
\be
iD_0 \hat{N} \sim 0 \,,
\ee
is used to eliminate additional terms containing $D_0$.  
The first term in $\hat{\cal L}^{(2)}$ defines the nucleon mass, 
${\hat{c}^{(2)}_1 } \equiv 1$. 
For the remaining terms,
${\hat{c}^{(2)}_2 }$
and 
${\hat{c}^{(2)}_3 }$
give the isoscalar and isovector magnetic moments, 
and 
${\hat{c}^{(2)}_4 }$
is a relativistic correction to the axial-vector coupling. 

At third order, the number of terms continues to proliferate.  Many 
of these are, at tree level, simply relativistic corrections to the
leading order terms that are summed automatically in the relativistic 
formalism.  
For simplicity in the present discussion, 
let us concentrate on those operators that do not involve the nucleon
spin, i.e., operators that can give rise to coherent interactions on adjacent 
nucleons.  
This case is particularly simple, and there is a unique operator up 
to isospin combinations, 
\begin{align}\label{lhat3}
\hat{\cal L}^{(3)} &= 
\hat{N}^\dagger \bigg(
 \hat{c}^{(3)}_1 \epsilon^{ijk} {\rm Tr}( A_i D_j D_k )
+ \hat{c}^{(3)}_2 \epsilon^{ijk} \tau^a {\rm Tr}( \tau^a A_i D_j D_k ) + \dots 
\bigg) \hat{N} \,. 
\end{align} 
These operators are mapped onto by the $\rho$ and $\omega$ exchange 
discussed earlier, 
The expansions (\ref{lhat1}),(\ref{lhat2}),(\ref{lhat3})
show explicitly that the operators parameterized by 
$\hat{c}^{(3)}_{1,2}$ in (\ref{lhat3}),  
and hence $c^{(3)}_{4,5}$ in (\ref{eq:expansion}), 
are the only direct interactions 
with a coherent coupling of one vector and one axial-vector field to the nucleon
through three-derivative order. 
Eqs.(\ref{nrlimit}) and Appendix~\ref{sec:app} below will verify
that iterations of ${\cal L}^{(1)}$ and ${\cal L}^{(2)}$
do not give rise to further coherent interactions. 
This justifies the statement made at the end of the previous section regarding
the unique property of the operator induced by the pCS terms, i.e.,
coherent coupling of vector and axial-vector fields to the nucleon.

\subsubsection{Low energy: integrating out the intermediate mass scale\label{sec:low}} 

Since the nucleon kinetic energy is
\be 
E_N - m_N \sim \bm{k}^2/2m_N \ll |\bm{k}| \,,
\ee
any exchange of energy of order $|\bm{k}|$ takes the nucleon 
far offshell.   
At very low energies we should integrate out such modes, making use
of the scale separation $m_N^2 \gg m_N |\bm{k}| \gg |\bm{k}|^2$. 
The scale $m_N^2$ was integrated out in the previous step, summarized by 
(\ref{nrexpand})  
(the same is accomplished, though not explicitly, 
by (\ref{rexpand}) in the relativistic formulation).

Contributions to the Compton-like scattering process depicted 
in Fig.~\ref{fig:compton}
begin at $\order(1/m_N)$.  Leading contributions involve 
a single insertion of $\hat{\cal L}^{(2)}$, 
while subleading contributions arise from either two 
insertions of $\hat{\cal L}^{(2)}$, or
a single insertion of $\hat{\cal L}^{(3)}$.    
Such combinations of $\hat{\cal L}^{(1)}$ 
and $\hat{\cal L}^{(2)}$ are described at very 
low energy by local contact
interactions corresponding to the 
amplitudes derived below in (\ref{nrlimit}) and (\ref{subleadingcompton}).  
In addition, there is the direct contribution from $\hat{\cal L}^{(3)}$, 
corresponding to (\ref{eq:omega_amp}) below. 
The competing energy scales for these different contributions 
are ${1/(m_N E)}$ for
the Compton-like process, versus $1/m_\rho^2 \sim 1/m_\omega^2$ 
for the meson exchange.

The nonrelativistic expansion provides an explicit scale separation that 
is insightful but not essential at tree level.  
Rather than pursue a more formal description of this low-energy effective
theory,
the relevant amplitudes are calculated in the following sections 
directly from the relativistic formulation (\ref{eq:expansion}).

\section{Low-energy constants and large-energy extrapolation \label{sec:model}} 

In order to proceed phenomenologically, the 
normalizations (Wilson coefficients or low-energy constants) 
for the relevant interactions of the lagrangian (\ref{eq:expansion}) must be fixed, 
and a reasonable model (form factors) must be specified
for extrapolating into the ${\rm GeV}$ energy range where the chiral
lagrangian is breaking down.  

It should be emphasized that in extrapolating to larger energy, we are leaving
the firm theoretical footing of the chiral lagrangian. 
The simplest form factors based on vector 
dominance are still predictions obtained at tree level from 
a well-defined effective lagrangian (including vector mesons), 
albeit with significant corrections from neglected loop effects.  
Introducing phenomenological form factors takes us further from a simple
effective lagrangian approach, 
but crudely accounts for effects of higher resonances, 
and provides a connection to higher energy
via perturbative scaling laws that must be satisfied when momentum invariants
are large.   Formal justification for this approach, in particular 
approximating amplitudes by 
tree level exchange of physical mesons, 
can be found in a large $N_c$ (number of colors) limit~\cite{Witten:1979kh}. 
Given the relatively modest extrapolations involved 
(e.g. up to $E_\nu \sim 1.5\,{\rm GeV}$), and the absence of a more 
controlled expansion scheme in this energy regime, 
the remainder of the paper proceeds without further apology.   

The rest of this section investigates 
the mechanisms of generalized Compton scattering; 
$t$-channel $\omega$ (and $\rho$, $\pi$) exchange; 
and $s$-channel $\Delta$ production.  
These mechanisms include the direct couplings of the electroweak gauge 
fields to nucleons, and incorporate effects of the dominant hadronic
resonances in each channel. 

\subsection{Compton scattering}

\begin{figure}
\begin{center}
\includegraphics[width=10pc, height=10pc]{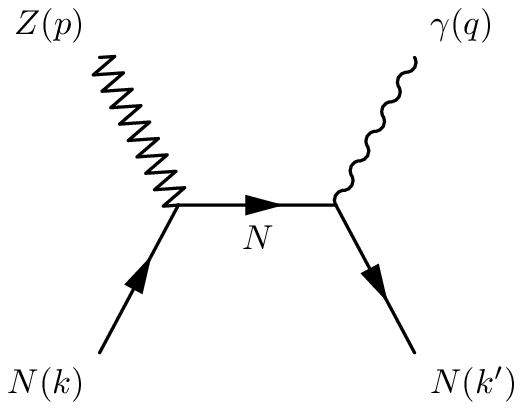}
\hspace{10mm}
\includegraphics[width=10pc, height=10pc]{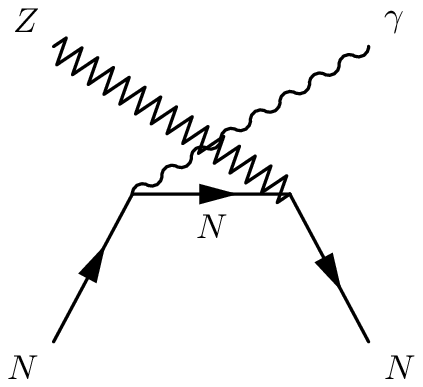}
\caption{\label{fig:compton}
Generalized compton scattering.
}
\end{center}
\end{figure}

Let us begin by examining the contributions to 
$\nu N \to \nu N \gamma$ mediated by an intermediate nucleon
as depicted in Fig.~\ref{fig:compton}.
These contributions will be referred to as ``Compton-like'' scattering 
where one of the photons is replaced by an (offshell) $Z$ boson.  
As discussed above, form factors for onshell nucleons are employed 
at the vertices to account for resonant structure 
in the appropriate channel. 

\subsubsection{Form factors}

The onshell matrix element of the weak neutral current
and electromagnetic current take the form
\begin{align}
\langle N(k')| J_{\rm NC}^\mu | N(k)\rangle 
&= 
{g_2\over 2 c_W} \bar{u}(k') \Gamma_{\rm NC}^\mu(k'-k) u(k) \,, \nl
\langle N(k')| J_{\rm em}^\mu | N(k)\rangle 
&= e\,\bar{u}(k') \Gamma_{\rm em}^\mu(k'-k) u(k) \,.
\end{align}
For the weak neutral current
\begin{align}\label{eq:weak} 
\Gamma_{\rm NC}^\mu(q) &= 
\gamma^\mu [ F^{1\,,\rm weak}_V(q^2) -F_A(q^2) \gamma_5 ] 
+ {i\over 2 m_N}  \sigma^{\mu\nu} q_\nu  F_V^{2\,,\rm weak}(q^2) 
+ {1\over m_N} F_P(q^2) q^\mu \gamma_5 
\,,
\end{align} 
and similarly, for the electromagnetic current: 
\begin{align} 
\Gamma_{\rm em}^\mu(q^2) &= \gamma^\mu 
F^{1\,,\rm em}_V(q^2) 
+ {i\over 2 m_N}  \sigma^{\mu\nu} q_\nu  F_V^{2\,,\rm em}(q^2) \,.
\end{align}
Enforcing time-reversal invariance ensures that 
$F_V^{1,2}(0)$, $F_A(0)$ and $F_P(0)$ are real
as expected from the effective lagrangian (\ref{eq:expansion}).  
Note that $F_P$ in (\ref{eq:weak}) is induced by 
pion exchange, and is not represented by a new operator in 
the theory~\cite{Bernard:1994wn}.
The contribution from $F_P$ vanishes when the current couples
to massless leptons (e.g., in the approximation of massless neutrinos), 
and will not be relevant here. 

Standard parameterizations in terms of the electric and magnetic 
form factors are used:~\cite{Llewellyn Smith:1971zm}
\begin{align}
G_E &\equiv F^1 + {q^2\over 4 m_N^2} F^2 \,, \nl
G_M &\equiv F^1 + F^2 \,,
\end{align} 
with 
\be\label{compff}
G_{E\,\rm proton} 
\approx {G_{M\,\rm proton}\over 1 + a_p} 
\approx {G_{M\,\rm neutron}\over a_n} 
\approx {1\over ( 1 - q^2/ 0.71\,{\rm GeV}^2 )^2}
\equiv F_D(q^2) \,.
\ee
In terms of the common overall dipole factor, the 
electromagnetic form factors of the proton are:
\begin{align}
F^{1\,,\rm em}_{V\,\rm proton}/F_D &= 1 - {q^2/4m_N^2\over 1-q^2/4m_N^2} a_p \,, \nl
F^{2\,,\rm em}_{V\,\rm proton}/F_D &= {a_p \over 1-q^2/4m_N^2} \,,
\end{align}
and for the neutron:
\begin{align}
F^{1\,,\rm em}_{V\,\rm neutron}/F_D &= - {q^2/4m_N^2\over 1-q^2/4m_N^2} a_n \,, \nl
F^{2\,,\rm em}_{V\,\rm neutron}/F_D &= {a_n \over 1-q^2/4m_N^2} \,.
\end{align}
Similarly, for the weak current, for the proton:
\begin{align}
F^{1\,,\rm weak}_{V\,\rm proton}/F_D &= 
\frac12-2s_W^2 
-{q^2/4m_N^2\over 1-q^2/4m_N^2}
\left[ \left(\frac12-2s_W^2\right)a_p - \frac12 a_n \right] 
\,, \nl
F^{2\,,\rm weak}_{V\,\rm proton}/F_D &= 
{1\over 1-q^2/4m_N^2} 
\left[ \left(\frac12-2s_W^2\right)a_p - \frac12 a_n \right] \,,
\end{align}
and for the neutron:
\begin{align} 
F^{1\,,\rm weak}_{V\,\rm neutron}/F_D &= 
-\frac12 
-{q^2/4m_N^2 \over 1-q^2/4m_N^2}
\left[ \left(\frac12-2s_W^2\right)a_n - \frac12 a_p \right] 
\,, \nl
F^{2\,,\rm weak}_{V\,\rm neutron}/F_D &= 
{1\over 1-q^2/4m_N^2} 
\left[ \left(\frac12-2s_W^2\right)a_n - \frac12 a_p \right] \,.
\end{align}
At $q^2=0$ the weak form factors reduce to 
$F^{1\,, \rm weak}_{V\, \rm proton}(0) \equiv C_{V\,\rm proton}=\frac12 - 2s_W^2$, 
$F^{1\,, \rm weak}_{V\, \rm neutron}(0) \equiv C_{V\,\rm neutron}=-\frac12$. 
Finally, a standard prescription is used for the axial-vector form factor:
\be\label{dipff}
F_A(q^2) = {F_A(0) \over (1 - q^2/m_A^2 )^2} \,,
\ee 
with axial mass parameter $m_A \approx 1.2\,{\rm GeV}$~\cite{:2007ru}.
For the normalization of the axial-vector coupling,
strange quark effects are ignored and the interaction is pure isovector:
\be
F_{A\,\rm proton}(0) \equiv C_{A\, \rm proton} = 
- F_{A, \rm neutron}(0)
\equiv - C_{A\, \rm neutron} 
= 1.26/2 \,.
\ee
The dipole form factors (\ref{compff}) and (\ref{dipff}) 
reproduce the correct perturbative scaling behavior at $q^2\to \infty$~\cite{Lepage:1980fj}.  

\subsubsection{Leading order cross section}

Consider the low energy limit, or equivalently the limit $m_N\to \infty$. 
The amplitude for $Z(p) + N(k) \to \gamma(q) + N(k')$ from the 
diagrams in Fig.~\ref{fig:compton} is: 
\begin{align}\label{nrlimit} 
  i{\cal M} = - { i e g_2 \over c_W}\, \chi^\dagger \bigg\{
  & 
  \epsilon^{(\gamma)*}_0 \epsilon^{(Z)}_0 
\bigg[ F^1 C_V \hat{\bm p}\cdot\hat{\bm q}  
- F^1 C_A  \hat{\bm q}\cdot \vec{\bm \sigma}  
\bigg]
\nl
+&
\epsilon^{(\gamma)*}_i \epsilon^{(Z)}_0 
\bigg[ F^1 C_V \hat{p}^i  
- F^1 C_A \sigma^i  
\bigg]
\nl
+&
\epsilon^{(\gamma)*}_0 \epsilon^{(Z)}_j 
\bigg[ F^1 C_V \hat{q}^j  
- F^1 C_A \hat{\bm p}\cdot \hat{\bm q} \sigma^j 
\bigg]
\nl
+&
\epsilon^{(\gamma)*}_i \epsilon^{(Z)}_j 
\bigg[ F^1 C_V \delta^{ij}  
- F^1 C_A\left( \hat{p}^i \sigma^j + 
\hat{q}^j\sigma^i - \delta^{ij}\hat{\bm q}\cdot{\bm\sigma}
\right)
\nl
& \qquad \qquad
- F^2 C_A (\hat{q}^j\sigma^i - \delta^{ij}\hat{\bm q}\cdot{\bm\sigma}) 
\bigg]
\bigg\} \chi \,.
\end{align}
When, as in reality, the $Z$ is virtual and connected to the neutrino 
line, the amplitude is given by the replacements
\begin{align}
{g_2 \over 2 c_W} \epsilon^{(Z)}_\mu &\to - {G_F\over \sqrt{2} }
\bar{\nu}(p') \gamma_\mu (1-\gamma_5) \nu(p) \,, \nl
\hat{p}^i &\to {1\over E_\gamma} (p^i - p^{\prime i} )\,,
\end{align}
where the reaction is $\nu(p)N(k)\to \nu(p')N(k')\gamma(q)$.  
Note that terms with $C_A$ involve spin-flip matrix elements, and hence 
the amplitudes for scattering on adjacent nuclei cannot add coherently.   
Appendix~\ref{sec:app} verifies this property also at the next derivative order. 
These computations complete the discussion after (\ref{lhat3}): 
the operators induced by the pCS terms 
give the only contributions through three derivative order 
that involve the axial component of the $Z$ boson, 
and that can be coherent on adjacent nucleons.     

Taking the $m_N\to\infty$ limit of the final state phase space, the cross
section for $\nu N \to \nu N \gamma$ arising from generalized 
Compton scattering becomes 
\begin{align}\label{comptonxs}
{d \sigma({\rm Compton}) \over de dx}  &=
{1\over \pi^2} {\alpha G_F^2 E^4 \over m_N^2} 
e(1-e) 
\bigg\{ \nl
&\hspace{-15mm} \quad 
F_1^2 C_V^2 \bigg[ {1\over e^2}\left(\frac12-\frac16 x^2 \right)
+{1\over e}\left(-\frac76+\frac56 x^2 \right) 
+ \frac43 -\frac23 x^2 -\frac23 e \bigg] 
\nl
&\hspace{-15mm} 
+ F_1^2 C_A^2\bigg[ {1\over e^2}\left( {17\over 6}-{11\over 6} x^2\right)
+{1\over e}\left(-{11\over 2}+{19\over 6}x^2 \right) 
+ 6 - 2x - \frac43 x^2 + e\left(-{10\over 3} + 2x\right)
\bigg] 
\nl
&\hspace{-15mm} 
+ F_1 F_2 C_A^2 \bigg[ (1-e)(4-2x) \bigg] 
\nl
&\hspace{-15mm}
+ F_2^2 C_A^2 \bigg[ 2(1-e) \bigg] \bigg\}\,. 
\end{align}
Here $x\equiv\cos\theta_\gamma$ and $e\equiv E_\gamma/E$, 
where $\theta_\gamma$ is the angle between the photon and the 
incoming neutrino, and $E_\gamma$, $E$ are the energies of the
photon and incoming neutrino.   
Note that there is a logarithmic singularity at $e\to 0$ in the
terms $F_1^2 C_V^2$ and $F_1^2 C_A^2$, corresponding to production of
very soft photons, i.e., bremsstrahlung corrections to
neutral current neutrino-nucleon scattering.  
For production of
photons above a fixed energy threshold, this infrared 
singularity does not pose a problem%
\footnote{We will not be concerned with extreme scale separations where 
logarithms $\log E/E_{\rm min}$ become large.}.  
   
\subsection{$t$-channel meson exchange} 

\begin{figure}
\begin{center}
\includegraphics[width=10pc, height=10pc]{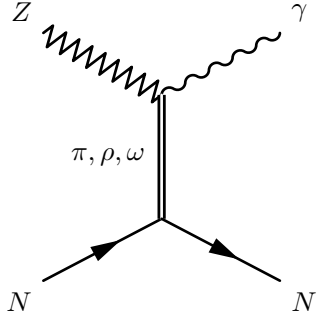}
\caption{\label{fig:meson}
Meson exchange contribution to $Z^* N \to \gamma N$. 
}
\end{center}
\end{figure}

Besides the diagrams in Fig.~\ref{fig:compton}, 
radiative neutrino scattering can take place via 
$t$ channel exchange of pseudoscalar and vector mesons, as depicted 
in Fig.~\ref{fig:meson}. 
Unlike Compton scattering, these contributions do not vanish 
in the zero-recoil limit.   

The relevant interactions at the upper vertex in this diagram are 
given by the lagrangian terms~\cite{HHH1}
\be
{\cal L} = {e g_2 \over 16\pi^2 c_W} \epsilon^{\mu\nu\rho\sigma} 
\bigg\{ 
\left(1-4s_W^2\right)  {\pi^0 \over f_\pi} \partial_\mu Z_\nu \partial_\rho A_\sigma 
- 3 g' \omega_\mu Z_\nu \partial_\rho A_\sigma 
- g \rho^{(0)}_\mu Z_\nu \partial_\rho A_\sigma \bigg\} \,. 
\ee
Note that there is no corresponding axial-vector meson exchange%
\footnote{By parity, an axial vector meson would necessarily couple to the photon 
and the vector $Z$, which cannot be combined with $\epsilon^{\mu\nu\rho\sigma}$ 
in a gauge invariant way.}.
Interactions of neutral mesons with nucleons take the form 
\be 
{\cal L} = g_{\pi NN} \partial_\mu \pi^0 
\bar{N} \tau^3 \gamma^\mu \gamma_5 N 
+ g_{\omega NN} \omega_\mu 
\bar{N} \gamma^\mu N 
+ g_{\rho NN} \rho^{0}_\mu 
\bar{N} \tau^3 \gamma^\mu N \,,
\ee
where $g_{\pi NN} \approx {g_A / 2 f_\pi}$ 
is the Goldberger-Treiman relation.  
From the identification of the nucleon as part of the isoscalar (baryon) and
isovector quark flavor currents, it follows that
\be\label{eq:coupling}
g_{\omega NN} \equiv g_\omega \sim \frac32 g' \,, \quad
g_{\rho NN} \sim \frac12 g \,, \quad 
\ee
where $g \sim g' \sim 6$.  
For extrapolating to higher energy, we adopt 
phenomenological form factors in the relevant channels.  At the 
$\omega$-$Z^0$-photon vertex in Fig.~\ref{fig:meson}
\be\label{nucff}
g' \to g'/(1- (p-p')^2/m_A^2) \,, 
\ee 
with $m_A\sim 1.2\,{\rm GeV}$ an axial-vector mass scale. 
This factor is induced by axial-vector meson interactions, and the 
single power of $[1-(p-p')^2/m_A^2]^{-1}$  
reproduces the correct scaling law for the $\omega$-$Z^0$-photon vertex,
in the limit $(p-p')^2 \to \infty$, 
$q^2=0$, $(k-k')^2 = {\rm constant}$;  
this is appropriate for an onshell 
photon, and small momentum transfer to the nucleus.  
At the $\omega$-$N$-$N$ vertex
\be 
g_{\omega NN} \to g_{\omega NN} / (1- (k-k')^2/\Lambda^2 ) \,,
\ee
where $\Lambda= 1.5\,{\rm GeV}$ is a phenomenological 
input~\cite{Machleidt:1987hj}.   
The combination of the $1-(k-k')^2/m_\omega^2$ factor 
from the $\omega$ propagator, and the additional $1-(k-k')^2/\Lambda^2$ factor in (\ref{nucff})
crudely represents the tower of higher-mass mesons exchanged in this channel.  
The details of these form factor models should not be taken too seriously.
Their main impact is to cut off amplitudes in regions of phase space that 
should not give large contributions to the cross sections.  The form 
factor parameters should be varied over a generous range to obtain 
reasonable error estimates in particular cases.  

To gauge the relative importance of the $\pi$, $\rho$ and $\omega$ 
contributions in Fig.~\ref{fig:meson}, consider the leading terms at
low energy.  
Using that $g_{\pi NN} \sim 1/f_\pi$, 
at very low energy the amplitude from pion exchange is parametrically of order 
${m_N E^3 / f_\pi^2 m_\pi^2}$ 
compared to Compton scattering.   
At energies large compared to the
pion mass, the pion is effectively massless, and 
the amplitude becomes of order
${m_N E / f_\pi^2}$ compared to Compton scattering.  
For the vector meson exchange, we have in contrast to (\ref{nrlimit}) the
amplitude
\begin{align}\label{eq:omega_amp}
i{\cal M} &\sim 
(\sqrt{2m_N})^2 {eg_2\over 16\pi^2 c_W  m_\omega^2} \chi^\dagger \chi 
(3 g' g_{\omega NN} \pm g g_{\rho NN} ) 
 \epsilon_i^{(\gamma)*} \epsilon_j^{(Z)} 
\epsilon_{ijk} q_k  \,,
\end{align}
where the $\pm$ refer to proton and neutron respectively. 
This demonstrates the claim made previously that 
the vector meson contributions are 
parametrically of order ${m_N E / m_\omega^2} \sim {m_N E/ m_\rho^2}$
compared to Compton scattering.  
Using (\ref{eq:coupling}) and $g' \sim g$, it follows that the 
$\omega$ contribution is approximately $3^2=9$ times larger in 
amplitude that the $\rho$ contribution.   
Contributions from states involving the strange quark 
are suppressed by their relatively small coupling to the nucleons.  
These facts, together with the suppression factor~\footnote{
The factor $1-4s_W^2$ results from tracing the vector component of the $Z$ 
with the electric charge matrix of the quarks and $\tau^3$ from the pion.
}
$1-4s_W^2\approx 0.08$ in the pion amplitude, 
indicate that $\omega$ gives the dominant meson-exchange
contribution to $\nu N \to \nu N \gamma$.   
This mechanism will compete with Compton scattering when $m_N E \gtrsim m_\omega^2$. 

For later use, the zero-recoil cross section for $\nu N \to \nu N \gamma$ 
resulting from $\omega$ exchange is (neglecting interference 
with other contributions)~\cite{HHH2}
\be\label{zero_omega}
{d\sigma{(\omega)} \over de dx} = {\alpha g_\omega^4 G_F^2 E^6 \over 16\pi^6 m_\omega^4}
e^3(1-e)^2 \,. 
\ee

\subsection{The $\Delta$ resonance}

\begin{figure}
\begin{center}
\includegraphics[width=10pc, height=10pc]{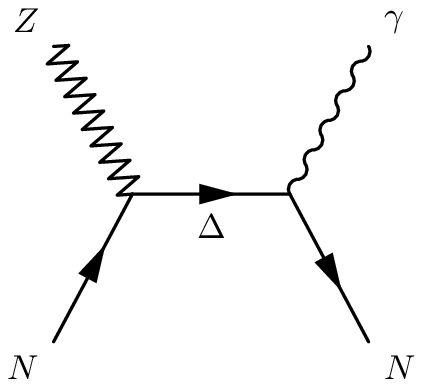}
\hspace{10mm}
\includegraphics[width=10pc, height=10pc]{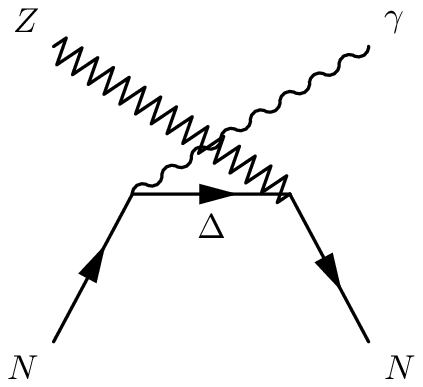}
\caption{\label{fig:delta}
Production of photons through the $\Delta$ resonance.
}
\end{center}
\end{figure}

At energies below $2\,{\rm GeV}$, $\Delta(1232)$
is the most prominent resonance appearing
in the $s$ (and $u$)
channels~\cite{Barish:1978pj,Radecky:1981fn,Kitagaki:1986ct,Kitagaki:1990vs}. 
We review here the salient features of including $\Delta$ as a field in 
our effective lagrangian, and derive matching conditions onto the low-energy
theory.  We will see that the leading effects at low energy are described 
by the same operator as for $t$-channel $\omega$ exchange.   

\subsubsection{Free Lagrangian}

A spin-$3/2$, isospin-$3/2$ particle such as $\Delta$
can be described by an isodoublet spinor field $\Delta^a_\mu$ 
carrying both isovector ($a=1..3$) and Lorentz ($\mu=0..3$) indices 
(as well as Dirac spinor indices and isodoublet indices which are suppressed)%
\footnote{
For a review see e.g. ~\cite{Hemmert:1997ye}.
}.
In this notation the constraint 
\be\label{iso}
\tau^a \Delta^a_\mu =0
\ee
must also be enforced 
to eliminate spurious isospin $1/2$ degrees of freedom%
\footnote{
A properly normalized basis is~\cite{Hemmert:1997ye}
$\Delta^{1} = (1/\sqrt{2}) [ \Delta^{++} -\Delta^0/\sqrt{3}, \, \Delta^+/\sqrt{3} - \Delta^- ]^T$, 
$\Delta^{2} = (i/\sqrt{2}) [ \Delta^{++} + \Delta^0/\sqrt{3}, \, \Delta^+/\sqrt{3} + \Delta^- ]^T$, 
$\Delta^3 = - (\sqrt{2/3}) [ \Delta^+, \, \Delta^0 ]$. 
}.
The free lagrangian may then be written 
\begin{align} 
\label{eq:delta_lagr}
{\cal L}_\Delta &= 
- \bar{\Delta}^a_\mu 
\bigg[ 
g^{\mu\nu} (i\slash{\partial} - m_\Delta ) - i(\gamma^\mu \partial^\nu + \gamma^\nu \partial^\mu)
+ i \gamma^\mu \slash{\partial} \gamma^\nu + m_\Delta \gamma^\mu \gamma^\nu 
\bigg] \Delta^a_\nu 
\nl
&= \bar{\Delta}^a_\mu 
\bigg[ 
\epsilon^{\mu\nu\alpha\beta} \gamma_5 \gamma_\alpha \partial_\beta 
+ i m_\Delta \sigma^{\mu\nu}
\bigg]
\Delta^a_\nu \,.
\end{align} 
The equations of motion from (\ref{eq:delta_lagr}) 
show that the free field satisfies 
\begin{align}\label{eom1}
\gamma^\mu \Delta^a_\mu &=0 \,, \\
(i\slash{\partial} - m_\Delta) \Delta^a_\mu &=0 \,,
\end{align}
which in turn imply that $\partial^\mu \Delta^a_\mu=0$. 
The Feynman rule for the $\Delta$ propagator ($\sim \langle \Delta^a_\mu \bar{\Delta}^b_\nu\rangle$) is 
\begin{multline}
\label{eq:delta_prop}
\frac23\left(\delta^{ab}-{i\over 2}\epsilon^{abc}\tau^c \right)  
{-i \over p^2 - m_\Delta^2 + i\epsilon} \times
\\
\times
\left[ 
\left( g_{\mu\nu} - {p_\mu p_\nu\over m_\Delta^2} \right) (\slash{p} + m_\Delta) 
+ \frac13 \left( \gamma_\mu + {p_\mu\over m_\Delta} \right) ( \slash{p} - m_\Delta ) 
\left(\gamma_\nu + {p_\nu \over m_\Delta} \right) \right] \,.
\end{multline}
Note that the unconventional prefactor results from the constraint (\ref{iso}). 
Other forms of the lagrangian may be obtained by a field 
redefinition $\delta \Delta^a_\mu \propto \gamma_\mu \gamma^\nu \Delta^a_\nu$. 
The resulting additional terms affect only the offshell behavior of the $\Delta$ 
and their effects 
are indistinguishable from the effects of 
local current-current interactions%
\footnote{  
$A=-1$ in the notation of \cite{Hemmert:1997ye}.
}.

\subsubsection{Interactions with nucleon and (axial-) vector fields}

To describe the processes pictured in Fig.~\ref{fig:delta} 
we must specify the interactions of $\Delta$ with nucleons 
and (axial-) vector fields.  
The lagrangian may be expanded as 
\be\label{eq:delta_int}
{\cal L}_{N\Delta} = {\cal L}_{N\Delta}^{(1)} 
+ {1\over m_N} {\cal L}_{N\Delta}^{(2)} 
+ {1\over m_N^2} {\cal L}_{N\Delta}^{(3)} + \dots 
 \,.
\ee
The leading operator is at one derivative order and includes couplings 
to pions and axial-vector fields: 
\begin{align}\label{eq:DNLO}
{\cal L}_{N\Delta}^{(1)} &= c_{N\Delta}^{(1)} 
\left[ \bar{\Delta}_\mu^a A^{a\, \mu} N  + h.c. \right] \,.
\end{align}
Normalization for the 
isovector components are defined as usual by 
$A_\mu=A_\mu^a\tau^a/2$.   
Coupling to vector fields begins at two-derivative order, 
\begin{align}\label{deltaff}
{\cal L}_{N\Delta}^{(2)} &=  
c^{(2)}_{N\Delta, 1} 
\left[ \bar{\Delta}_{\mu}^a 
F^{a\, \mu\nu} \gamma_\nu i\gamma_5
N  + h.c. \right]  + \dots \,, \nl
{\cal L}_{N\Delta}^{(3)} &= 
{c^{(3)}_{N\Delta, 1}} 
\left[
\bar{\Delta}_{\mu}^a 
iD_\nu ( F^{a\, \mu\nu} i\gamma_5 N )
+ h.c. \right]
+
{c^{(3)}_{N\Delta, 2}} 
\left[
\bar{\Delta}_{\mu}^a 
F^{a\, \mu\nu} iD_\nu 
i\gamma_5 N + h.c.\right] + \dots  \,,
\end{align}
where $F_{\mu\nu} \equiv i[D_\mu, D_\nu]$.  
Although they are naively of lower order (time derivatives acting on $N$ and $\Delta$), 
the operators with $c^{(3)}_{N\Delta, 1}$ and $c^{(3)}_{N\Delta, 2}$ 
are in fact 
power suppressed with respect to the operator with $c^{(2)}_{N\Delta, 1}$.
This can be seen by expanding on an explicit basis%
\footnote{
Note that 
the operators with $c^{(3)}_{N\Delta, 2}$ and $c^{(3)}_{N\Delta, 3}$ 
yield identical matrix elements for onshell massless vector fields such as the photon.  
}. 
For convenience in the phenomenological discussion these operators will be retained, 
although they have a relatively minor impact numerically.  
Similar power-suppressed terms in (\ref{deltaff}) involving 
the axial-vector field have been ignored.

\subsubsection{Normalization  and form factors}

The coefficients appearing in (\ref{eq:DNLO}), (\ref{deltaff}) can be 
determined from electro- or neutrino-production measurements 
$e N \to e \Delta$, $\nu N \to \ell_\nu \Delta$.  
The following default values are adopted:
\begin{align}\label{dcoeffs}
c^{(1)}_{N\Delta} &= {\sqrt{3 \over 2} } C^A_5 \approx 1.47 \,, \nl
c^{(2)}_{N\Delta,1} &= {-\sqrt{3 \over 2} } C^V_3 \approx -2.45 \,, \nl
c^{(3)}_{N\Delta,1} &= {-\sqrt{3 \over 2} } C^V_4 \approx 1.87 \,, \nl
c^{(3)}_{N\Delta,2} &= {-\sqrt{3 \over 2} } C^V_5 \approx 0 \,.
\end{align}
where $C^A_5 = 1.2$, $C^V_3=2.0$ and 
for simplicity the magnetic dominance approximation is employed: 
$C^V_4 = -(m_N/m_\Delta)C^V_3$, $C^V_5=0$. 
Within the relevant level of precision, these values reproduce resonance 
production data~\cite{Lalakulich:2005cs,Lalakulich:2006sw}. 
In extrapolating the chiral lagrangian to larger energy, we adopt 
phenomenological form factors
\begin{align}\label{dipole}
C^A_5 &\to C^A_5/(1-q^2/m_A^2)^2 \,, \nl
C^V_i &\to C^V_i/(1-q^2/m_V^2)^2 \,,
\end{align}
with $m_A\approx 1.0\,{\rm GeV}$ and $m_V\approx 0.8\,{\rm GeV}$.   
Corrections to dipole behavior of the form factors is neglected in the 
limited energy range under consideration.

\subsubsection{Induced interactions and influence of off-shell parameters}

With the vertices from (\ref{eq:delta_int}) and propagator from (\ref{eq:delta_prop})
it is straightforward to read off the induced interactions when $\Delta$ 
is integrated out of the theory.  
In particular, the leading term involving one 
axial-vector field and one vector field must take the general form given in (\ref{eq:expansion}).
Recall that only $c^{(3)}_{4,5}$ are relevant to neutral electroweak fields, 
and that $c^{(3)}_5$ involves the isoscalar component of the vector field, 
whereas only the isovector component can couple $N$ and $\Delta$. 
Thus only $c^{(3)}_4$ appears in the matching, and
an explicit calculation yields
\be\label{eq:delta_induce}
c^{(3)}_4(\Delta) = -\frac49 c_{N\Delta}^{(1)} c_{N\Delta,1}^{(2)} {m_N \over m_\Delta - m_N} 
\approx 5.2 \,. 
\ee
The sign of $c_4^{(3)}(\Delta)$ is fixed by the relative sign of $C^A_5$ and $C^V_3$ in 
(\ref{dcoeffs}), which in turn is confirmed phenomenologically 
by the larger $\nu$ versus $\bar{\nu}$ cross section
for $\Delta$ production.  

The matching condition 
(\ref{eq:delta_induce}) implies a large effect of the $s$-channel $\Delta$
besides the $t$-channel $\omega$.   
It is important to understand the robustness of this effect. 
Note that 
the result is affected by offshell modifications of the $\Delta$. 
For instance, to the leading lagrangian (\ref{eq:DNLO}) we can add terms such as
\be\label{eq:z}
\bar{\Delta}_\mu^a A^{a\,, \mu} N  
\to \bar{\Delta}_\mu^a (g^{\mu\nu} + z \gamma^\mu\gamma^\nu) A^{a}_{\nu} N \,.
\ee
The term involving $z$ does not affect on-shell properties, as seen 
by (\ref{eom1}), or in Feynman diagram language by explicitly 
contracting $\gamma_\mu$ with the propagator (\ref{eq:delta_prop}).   
This extra term is thus not constrained by onshell production measurements (apart from 
details in the lineshape).  
However, the perturbation (\ref{eq:z}) does affect offshell properties; in 
particular,  the new tree level matching condition becomes 
\be\label{eq:zdelta}
c^{(3)}_4(\Delta) \sim  
 -\frac49 c_{N\Delta}^{(1)} c_{N\Delta,1}^{(2)}
{ m_N^2 \over m_\Delta^2 - m_N^2} \left( { m_N+ m_\Delta \over m_N}  
+ z { m_\Delta^2- m_N^2 \over m_N m_\Delta } \right) \,.
\ee

Although at first sight it would appear that no prediction is possible without knowledge of $z$,
we do retain predictive power in the limit
\be\label{small}
m_\Delta, m_N \gg m_\Delta - m_N  \,. 
\ee
This is because amplitudes involving $z$ necessarily involve a factor that 
vanishes onshell, compensating the propagator singularity. 
Schematically, 
\be
{1\over p^2 -m_\Delta^2} \times (p^2 - m_\Delta^2)  \sim 1\,,
\ee
where $p$ is the momentum of an offshell $\Delta$.  
Such amplitudes correspond to local current-current interactions in the theory before
integrating out $\Delta$.      
In the low-energy context, this observation translates into the 
absence of new terms enhanced by factors of $(m_\Delta^2 -m_N^2)^{-1}$.  
Returning to (\ref{eq:zdelta}), 
we see that (\ref{eq:delta_induce}) gives the leading term containing 
an enhancement factor $m_N/(m_\Delta-m_N) \sim 3.2$.
For completeness, we note that in the same limit 
the remaining coefficients in 
the lagrangian (\ref{eq:expansion}) are 
\be
c^{(3)}_6 (\Delta)= - c^{(3)}_7(\Delta) = c^{(3)}_4(\Delta) \,. 
\ee
These relations can be seen easily 
from the structure of the propagator (\ref{eq:delta_prop}) 
between nonrelativistic nucleon states in the limit $(m_\Delta-m_N)/m_N\to 0$. 

In later applications, onshell production of $\Delta$ 
in Fig.~\ref{fig:delta} is described by modifying the 
propagator according to
\be
{1\over p^2 - m_\Delta^2} \to 
{1\over p^2 - m_\Delta^2 + i m_\Delta \Gamma_\Delta} \,,
\ee
where $\Gamma_\Delta \approx 120\,{\rm MeV}$. 
As a further refinement, the width can be assigned a 
dependence on energy determined by the dominant $N\pi$ 
decay mode: 
\be\label{width}
\Gamma_\Delta \to \Gamma_\Delta \left( p(W) \over p(m_\Delta) \right)^3 \,. 
\ee
Here $p$ is the $3$-momentum of the pion in the $\Delta$ rest frame: 
\be
p(W) = {1\over 2W}\sqrt{ (W^2-m_N^2-m_\pi^2)^2 - 4m_N^2m_\pi^2 } \,,
\ee
and the constraint $W\ge m_N+m_\pi$ is enforced on the 
invariant mass of the (offshell) $\Delta$. 

\subsubsection{Related pion processes and large $N_c$}

To judge the accuracy of the prediction (\ref{eq:delta_induce}), it is useful
to apply the same expansion to situations where the answer is relatively 
well known.   
For this purpose, and also for later comparison to 
pion production, let us consider the corrections to ${\cal L}^{(2)}$ in 
(\ref{eq:expansion}) 
that contain two axial-vector fields.   These terms may be written%
\footnote{
It is common to expand in terms of 
$F^+_{\mu\nu} \equiv 2i [D_\mu,D_\nu] -(i/2) [A_\mu,A_\nu]$ 
and $[A_\mu,A_\nu]$~\cite{Gasser:1987rb,Fettes:1998ud}. 
The operator $F^+_{\mu\nu}$ mixes axial-vector and vector 
components, but is convenient in some applications (e.g. pion scattering)
since it involves at least one external field.  
We choose instead to work with $[D_\mu, D_\nu]$ and $[A_\mu,A_\nu]$, keeping the 
distinction between vector and axial-vector explicit.   
The difference is irrelevant for neutral fields where $[A_\mu,A_\nu]$ vanishes.  
}
\be\label{eq:piN}
\Delta {\cal L}^{(2)} 
= \bar{N} \bigg\{ 
c^{(2)}_3 {\rm Tr}( A_\mu A^\mu)  
+ 
c^{(2)}_4 
\frac{1}{4m_N^2}\left[
{\rm Tr} (A_\mu A_\nu) \{ iD_\mu, iD_\nu \} 
+ h.c. \right]
+ \frac{i}{2} c^{(2)}_5 \sigma^{\mu\nu} [A_\mu, A_\nu] 
\bigg\} N \,. 
\ee
It is straightforward to compute, in analogy to (\ref{eq:delta_induce}), 
\be\label{eq:piNc}
c^{(2)}_3(\Delta) = -c^{(2)}_4(\Delta) = -2 c^{(2)}_5(\Delta) 
= - \frac89  
[ c^{(1)}_{N\Delta} ]^2 {m_N\over m_\Delta - m_N} \approx -6.2  \,. 
\ee
where we keep only the leading term in $m_N/(m_\Delta-m_N)$.  
Similar relations have been derived by Bernard et.al.~\cite{Bernard:1996gq};
they have included additional $N^*$ and $\sigma$ resonances, finding that
the $\Delta$ appears to give a dominant contribution to the matching. 
Experimental values from low-energy pion-nucleus scattering are~\cite{Fettes:1998ud}
\begin{align}
c^{(2)}_3 &= [-5.9(1) ] {\rm GeV}^{-1} \times  (2 m_N) =-11.1(2) \,, \nl
-c^{(2)}_4 &=- [ 3.2(2) ] {\rm GeV}^{-1}\times (2 m_N) = -6.2(4) \,, \nl
-2 c^{(2)}_5 &= -2 [ 3.47(5) ] {\rm GeV}^{-1}\times (2 m_N) = -13.0(2) \,. 
\end{align}
Dominance of $\Delta$ in this channel, and the limit of small $m_\Delta-m_N$, 
predicts the correct sign and approximate magnitude of
these low energy constants.  This lends support to taking seriously 
the large value indicated by (\ref{eq:delta_induce}).  

We can formalize the expansion in $(m_\Delta-m_N)/m_N$ by noticing that
this quantity is $\order(1/N_c^2)$ in the $1/N_c$ expansion.  
We can further 
make use of the large-$N_c$ relations $g_{\pi N\Delta} = \frac32 g_{\pi NN}$, 
$\mu_{N\Delta} = (\mu_p - \mu_n)/\sqrt{2}$~\cite{Adkins:1983ya}, 
which translate to  
\begin{align}\label{largeNcoeffs}
c^{(1)}_{N\Delta} &= {3\over 2\sqrt{2}} g_A = 1.3\,, \nl
c^{(2)}_{N\Delta, 1} &= - {3\over 4\sqrt{2}} (1 + a_p - a_n) = -2.5\,.
\end{align} 
These values are in good agreement with the phenomenological values 
quoted above in (\ref{dcoeffs}), although the remarkable precision is 
perhaps fortuitous.   
In terms of low-energy observables, the relation 
\be\label{pcsN}
c^{(3)}_4(\Delta) \approx { g_A \over 4} { 1+a_p - a_n \over m_\Delta/m_N - 1} = 4.7 \,,
\ee
is thus valid to leading order in $1/N_c$.
The relative sign in (\ref{largeNcoeffs}), and hence the overall sign in (\ref{pcsN}),
is confirmed by noticing that in the nonrelativistic and large $N_c$ limits, 
axial-vector and vector fields couple to the isovector magnetic moment operator 
for both $N$ and $\Delta$ 
proportional to $g_A A^{3\, i} + (1+a_p-a_n)/(2m_N) \epsilon^{ijk}\partial^j V^{3\, k}$, 
where $A^3_\mu$ and $V^3_\mu$ are the $a=3$ isospin components.

At fixed $N_c$ (e.g. $N_c=3$!), the counting rules
provide an understanding of the relative size of various contributions.  
It is amusing that 
(\ref{eq:piN}) and (\ref{eq:piNc}) result in an apparent 
violation of the $1/N_c$ counting rule stating that the amplitude 
for $\pi N\to \pi N$ scattering at fixed pion energy 
behaves as $(N_c)^0$ at large $N_c$%
\footnote{
Consider, e.g., the scalar isoscalar channel.  The apparent violation here
is in addition to that encountered in other channels when only nucleon 
intermediate states are considered; see e.g. the second reference of \cite{Witten:1979kh}.  
}. 
In this formal limit, the difference $m_\Delta-m_N$ would vanish, 
and the propagator scales as $( E_\pi m_N )^{-1} \sim N_c^{-1}$; after a cancellation 
between $s$ and $u$ channel diagrams, the usual scaling is reinstated.   
Thus, while there is no contradiction with the formal $N_c\to\infty$ limit, 
the large values of certain coefficients in the chiral lagrangian can be understood
as ``color enhancements''.   
A similar phenomenon appears in the coupling (\ref{pcsN}), where $c^{(3)}_4(\Delta)/m_N^2 \sim N_c^3$. 
Again, in the formal limit $m_\Delta - m_N \to 0$, the usual scaling is recovered. 
For comparison, $c^{(3)}_4(\omega)/m_N^2 \sim N_c^1$ has ``normal'' counting, 
and $c^{(3)}_5(\rho)/m_N^2 \sim N_c^{-1}$ has a suppressed counting.   
These counting rules result from $g_A\sim N_c$, $f_\pi \sim \sqrt{N_c}$, $m_\Delta - m_N \sim N_c^{-1}$, 
$m_N \sim N_c$, $g, g' \sim N_c^{-1/2}$, $g_{\omega NN} \sim N_c g'$ and $g_{\rho NN} \sim g$. 

The coefficient $c^{(3)}_4$ corresponds to a linear combination of 
terms studied in the related process $\gamma N \to \pi^0 N$~\cite{Bernard:1994gm}.   
The present analysis justifies the hierarchy of $\Delta$, $\omega$ and $\rho$ 
contributions found there in terms of a $1/N_c$ expansion%
\footnote{
After performing a field redefinition, the results of \cite{Ecker:1994pi} 
indicate that $c^{(3)}_4$ is not renormalized by pion loops.  
}.   
The $m_\Delta-m_N\to 0$ limit for
low-energy interactions of the nucleon and $\Delta$ with pions and gauge fields 
is naturally described with a solitonic (Skyrmion) representation
of the baryons.  
This will be discussed further in Section~\ref{sec:coh} below,
in relation to coherence effects.  

\section{Single nucleon phenomenology \label{sec:pheno}}

\begin{figure}
\begin{center}
\includegraphics[width=20pc, height=20pc]{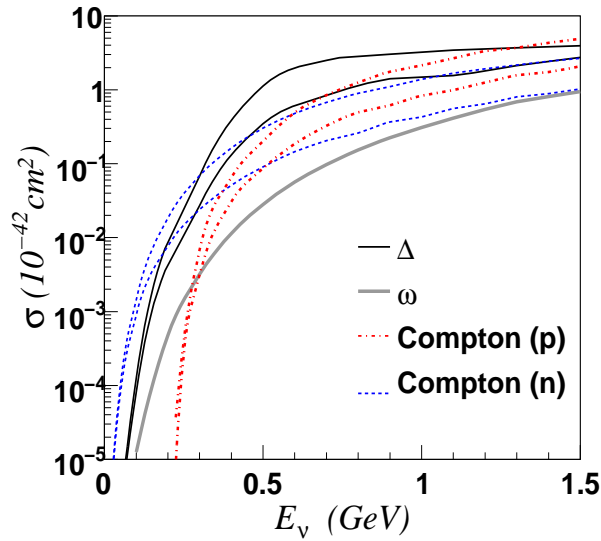}
\caption{
\label{fig:full}
Cross sections including recoil and form factors for 
$\nu N \to \nu N \gamma$ and $\bar{\nu} N \to \bar{\nu} N \gamma$. 
For the proton, a cut 
$E\ge 200\,{\rm MeV}$ is applied to the photon energy.
The $\omega$ contribution uses effective coupling $\frac32 g' = g_{\omega NN} = 10$.
The $\omega$ and $\Delta$ cross sections are identical for proton and neutron. 
For each of the Compton(proton), Compton(neutron) and $\Delta$ 
contributions, there are two curves, with the 
upper (lower) representing $\nu$ ($\bar{\nu}$).   
}
\end{center}
\end{figure}

\begin{figure}
\begin{center}
\includegraphics[width=12pc, height=12pc]{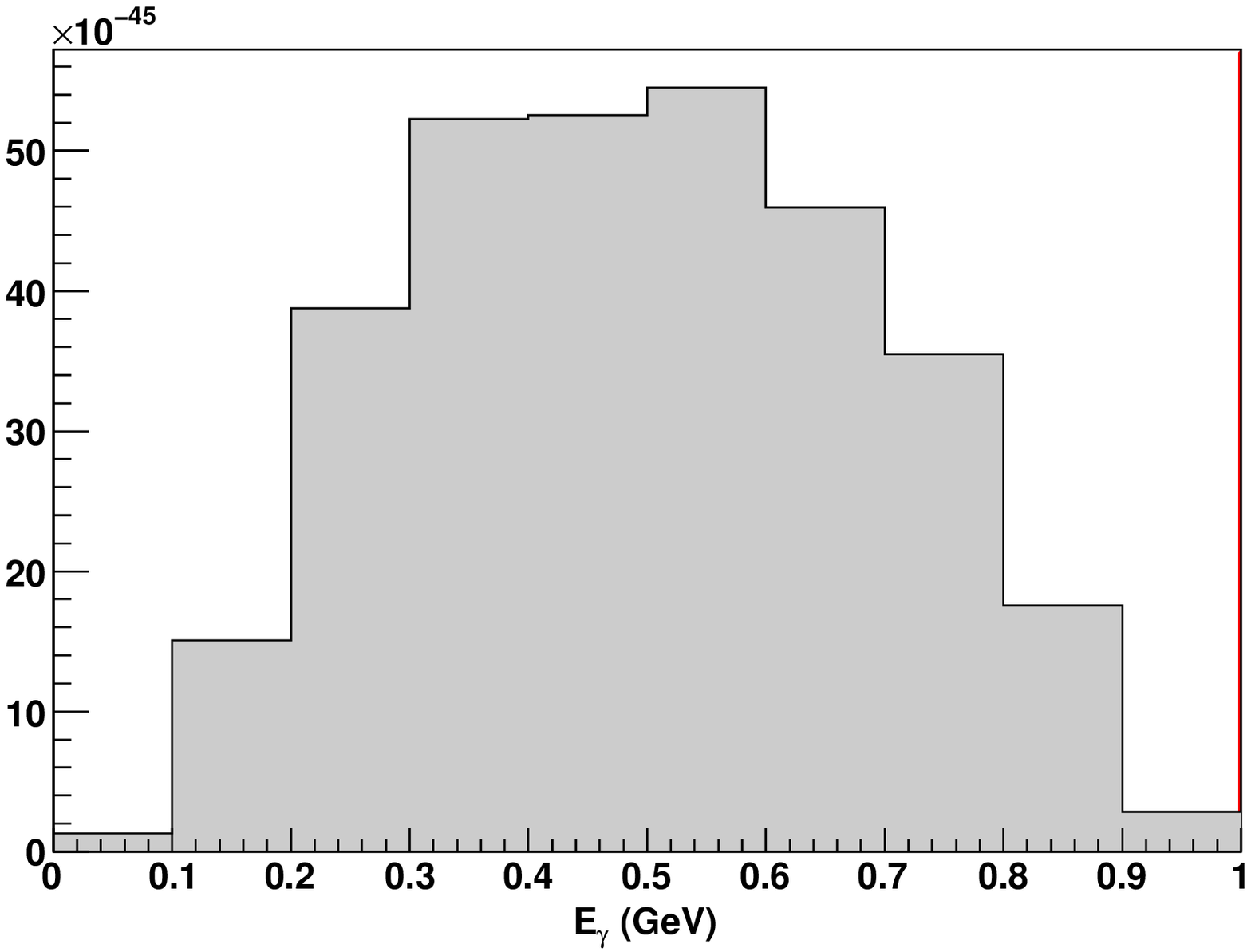}
\includegraphics[width=12pc, height=12pc]{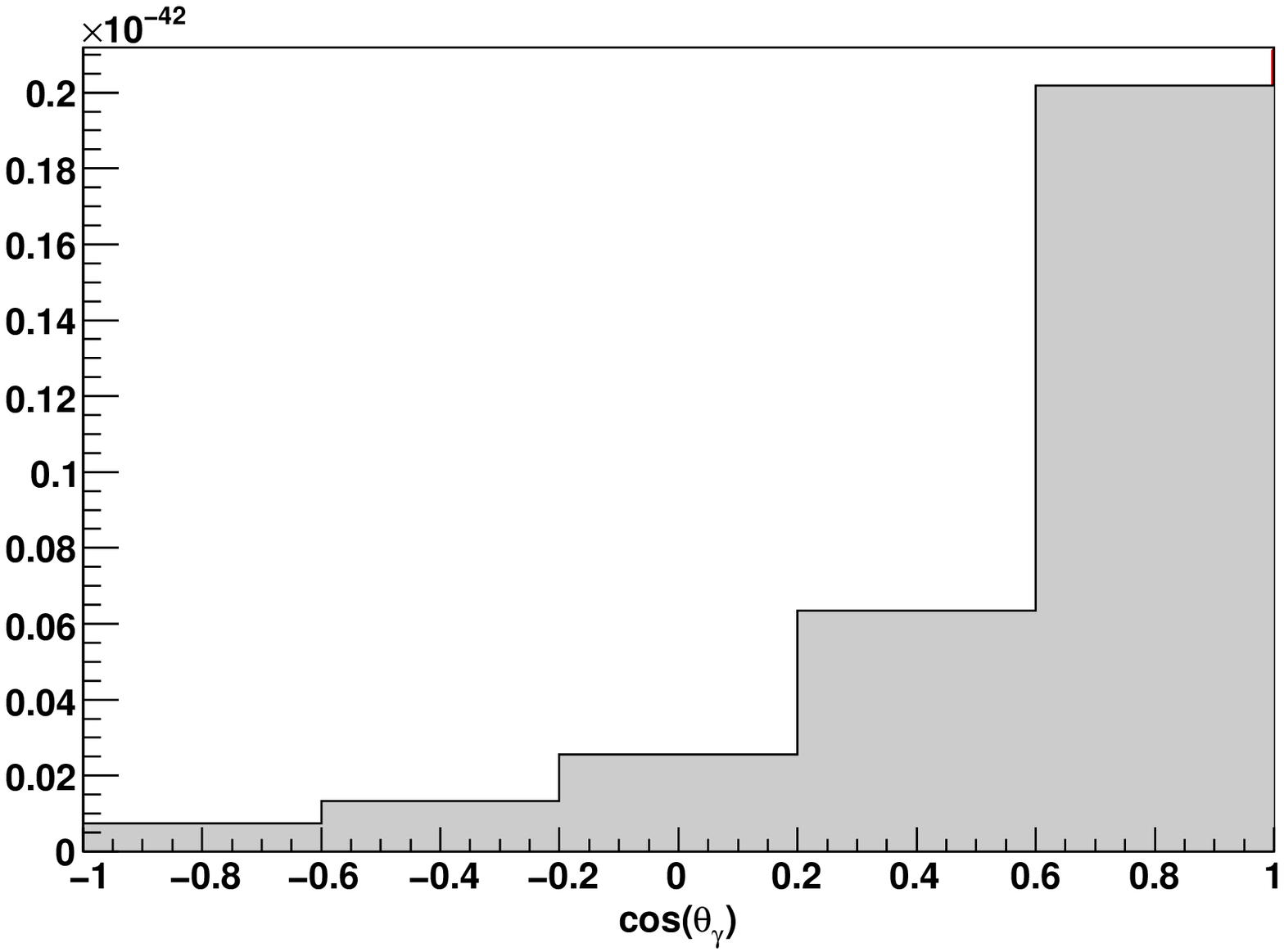}
\includegraphics[width=12pc, height=12pc]{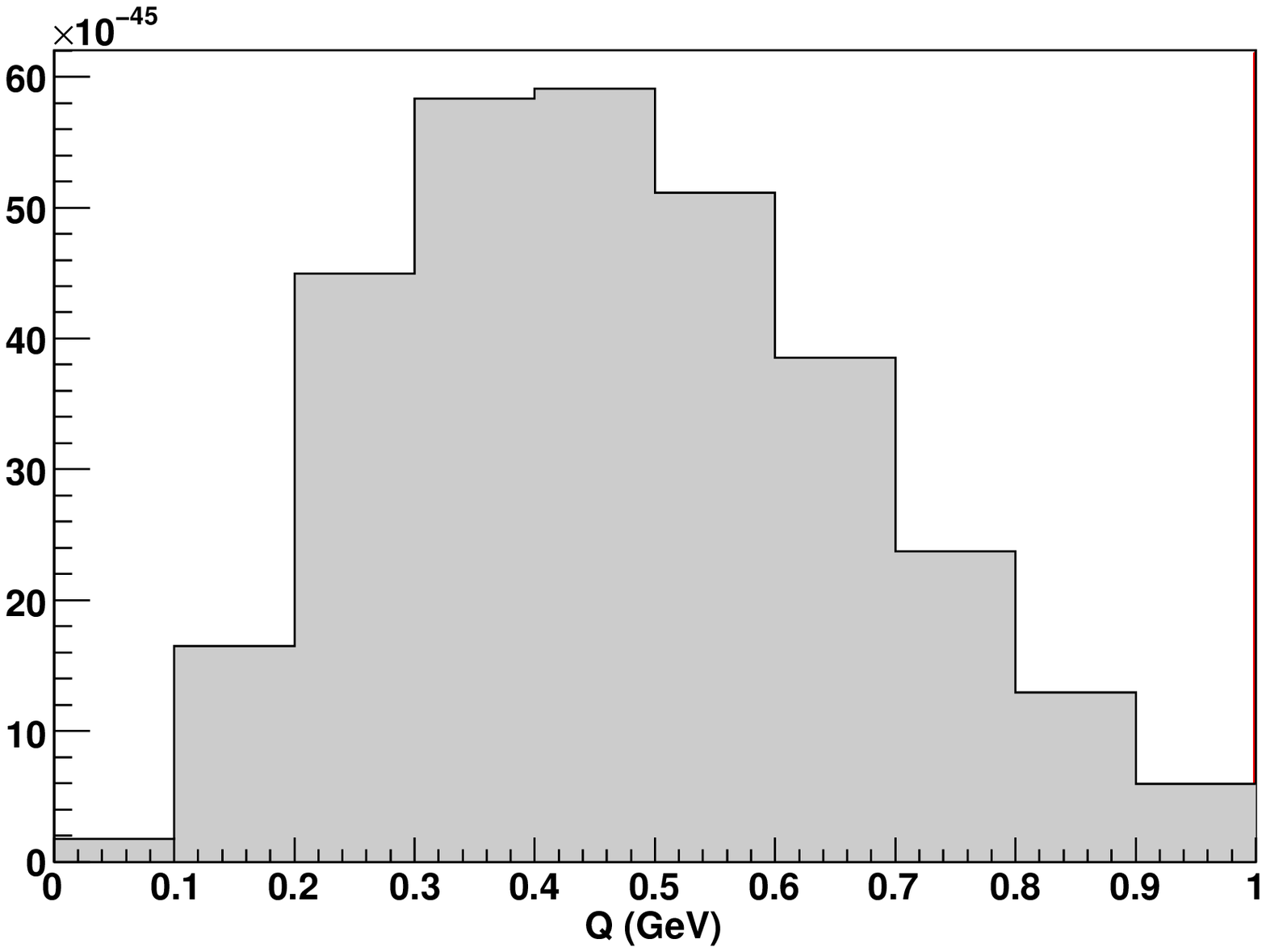}
\includegraphics[width=12pc, height=12pc]{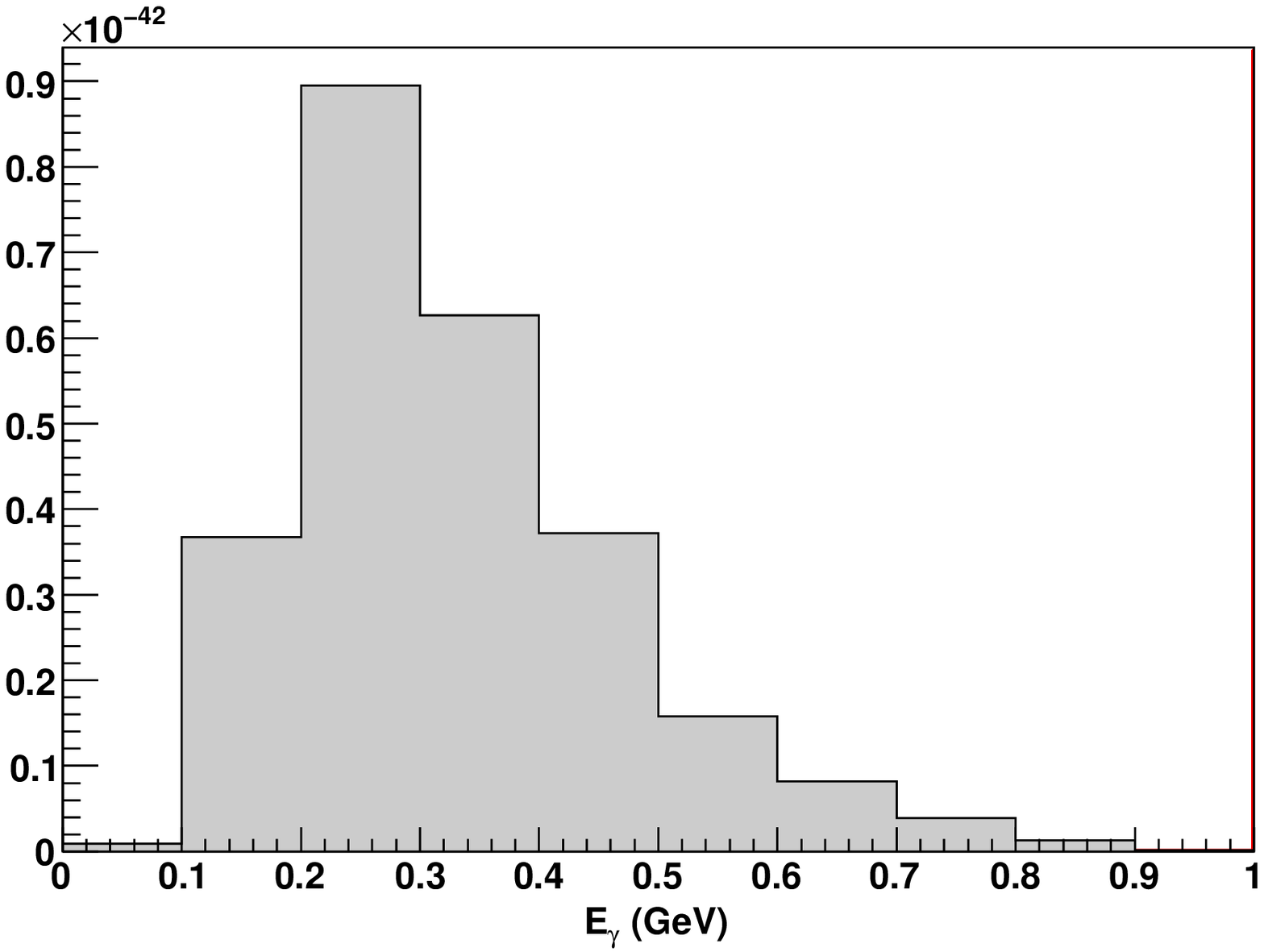}
\includegraphics[width=12pc, height=12pc]{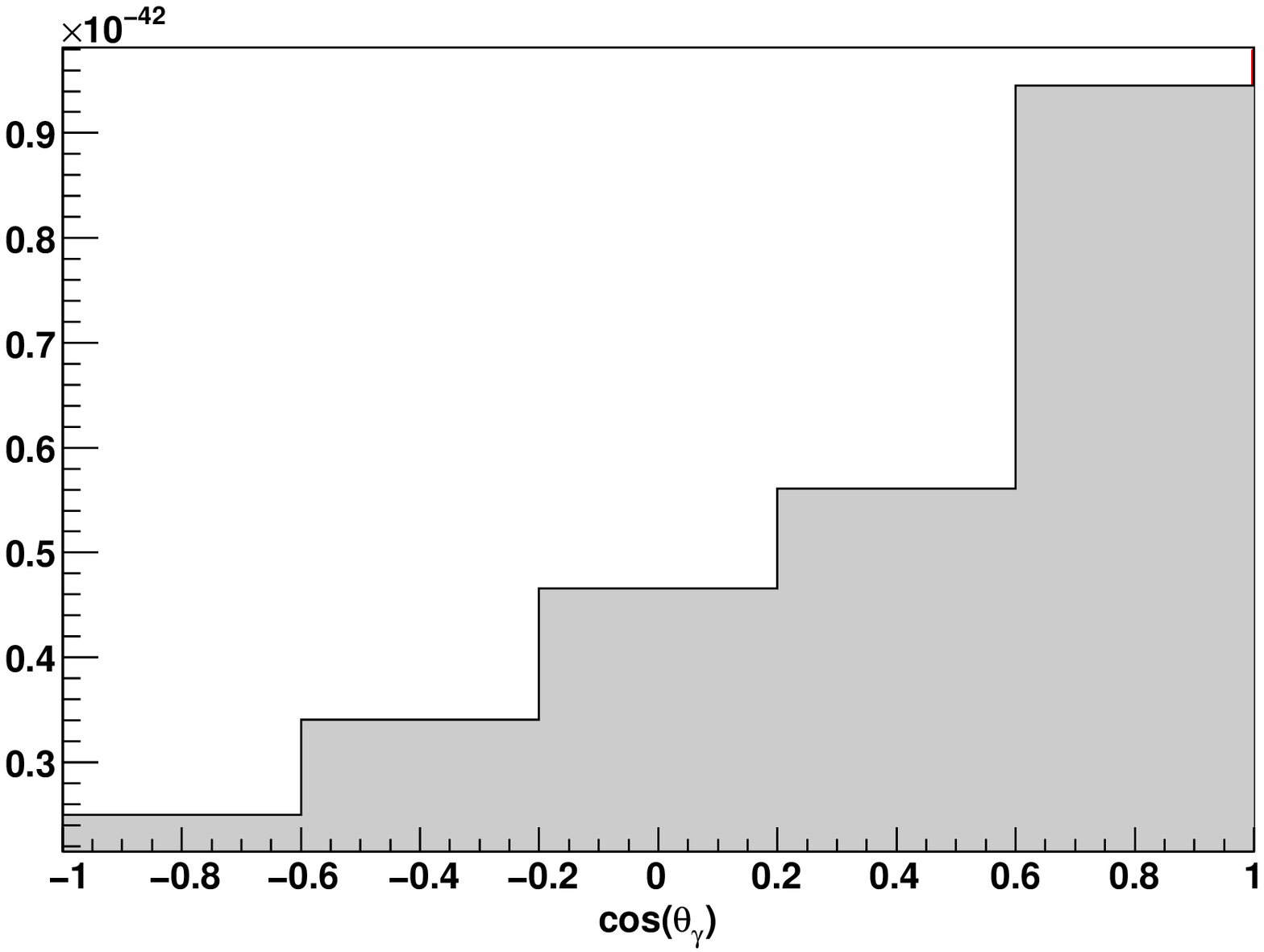}
\includegraphics[width=12pc, height=12pc]{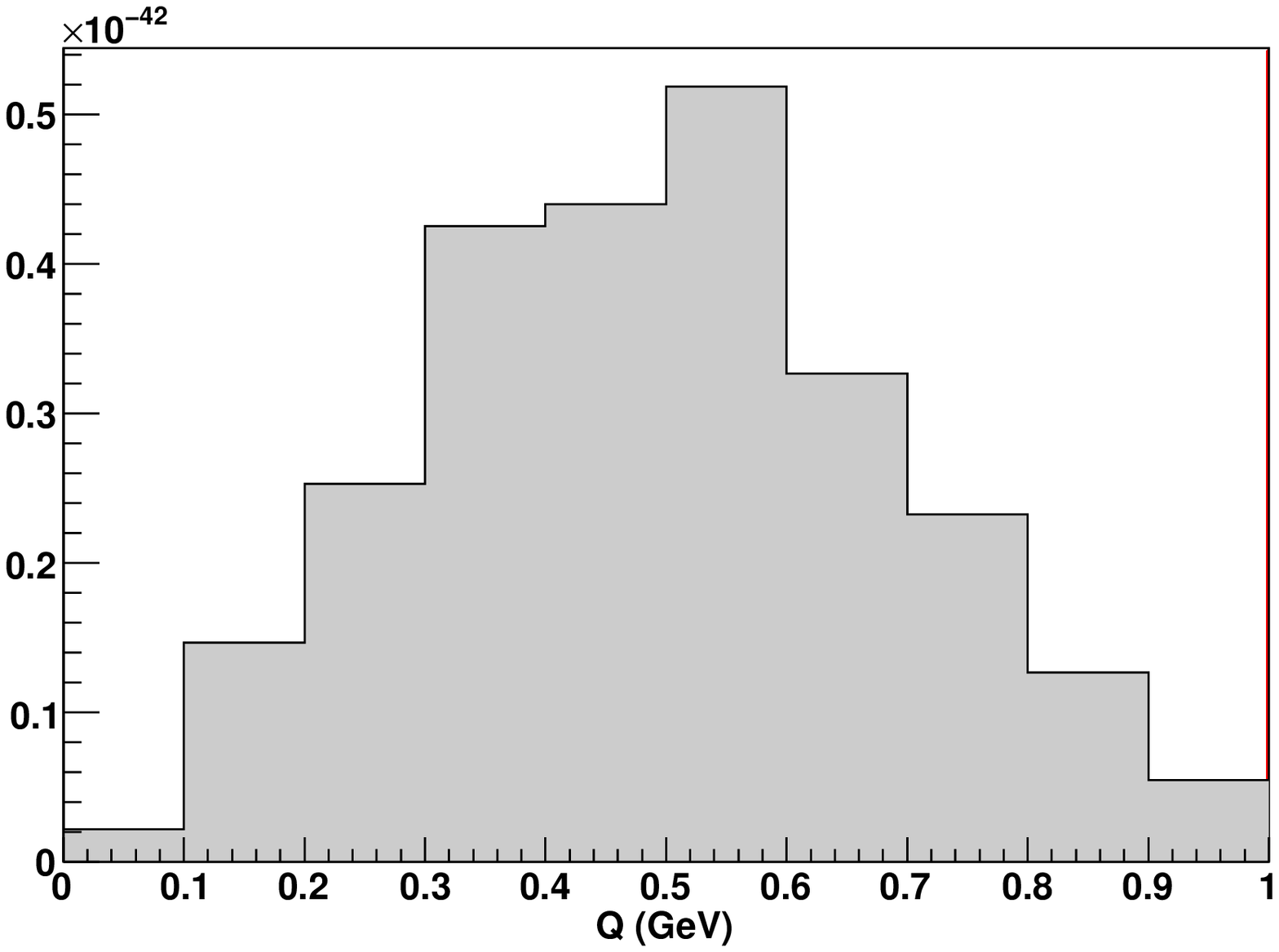}
\includegraphics[width=12pc, height=12pc]{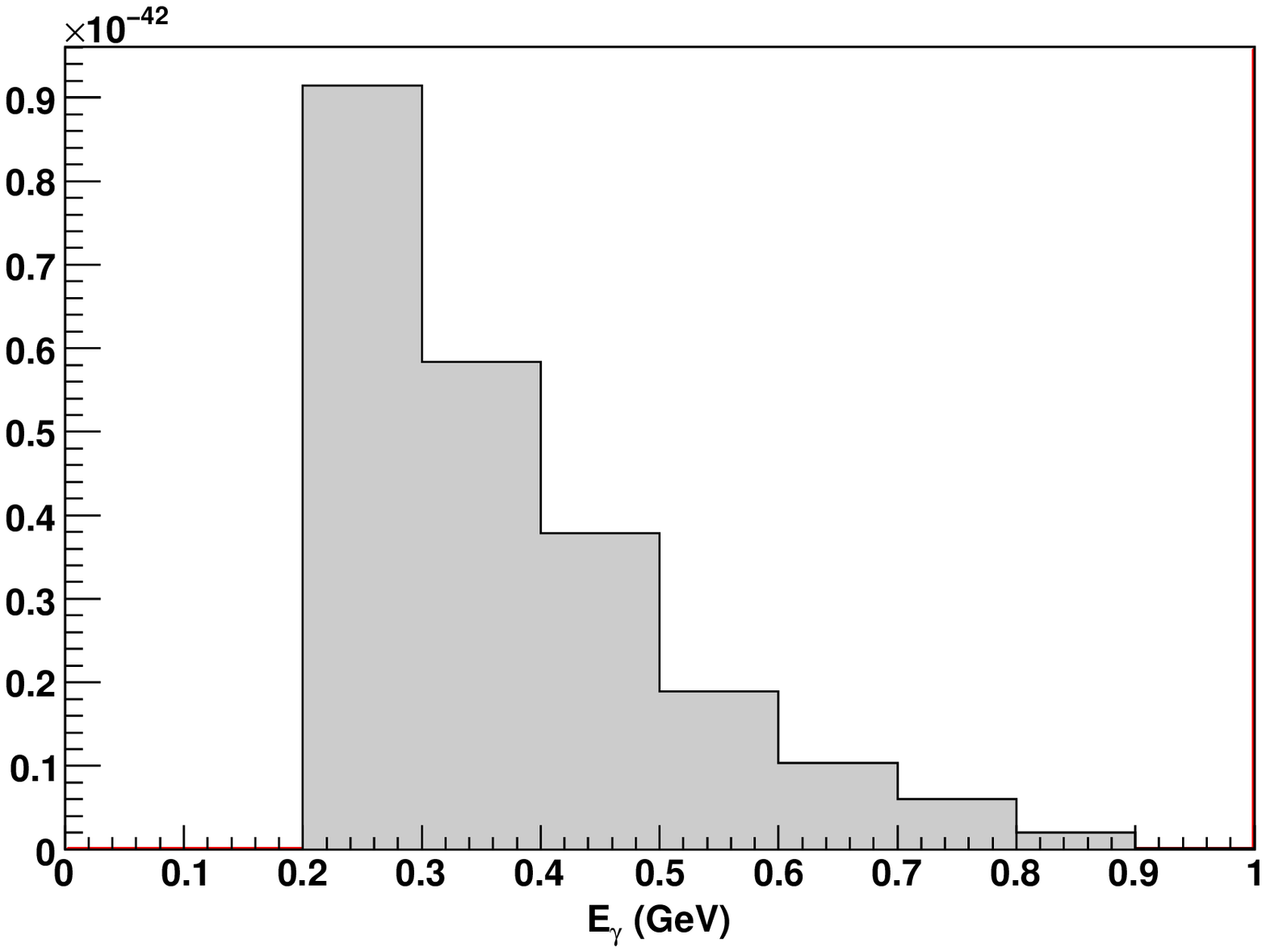}
\includegraphics[width=12pc, height=12pc]{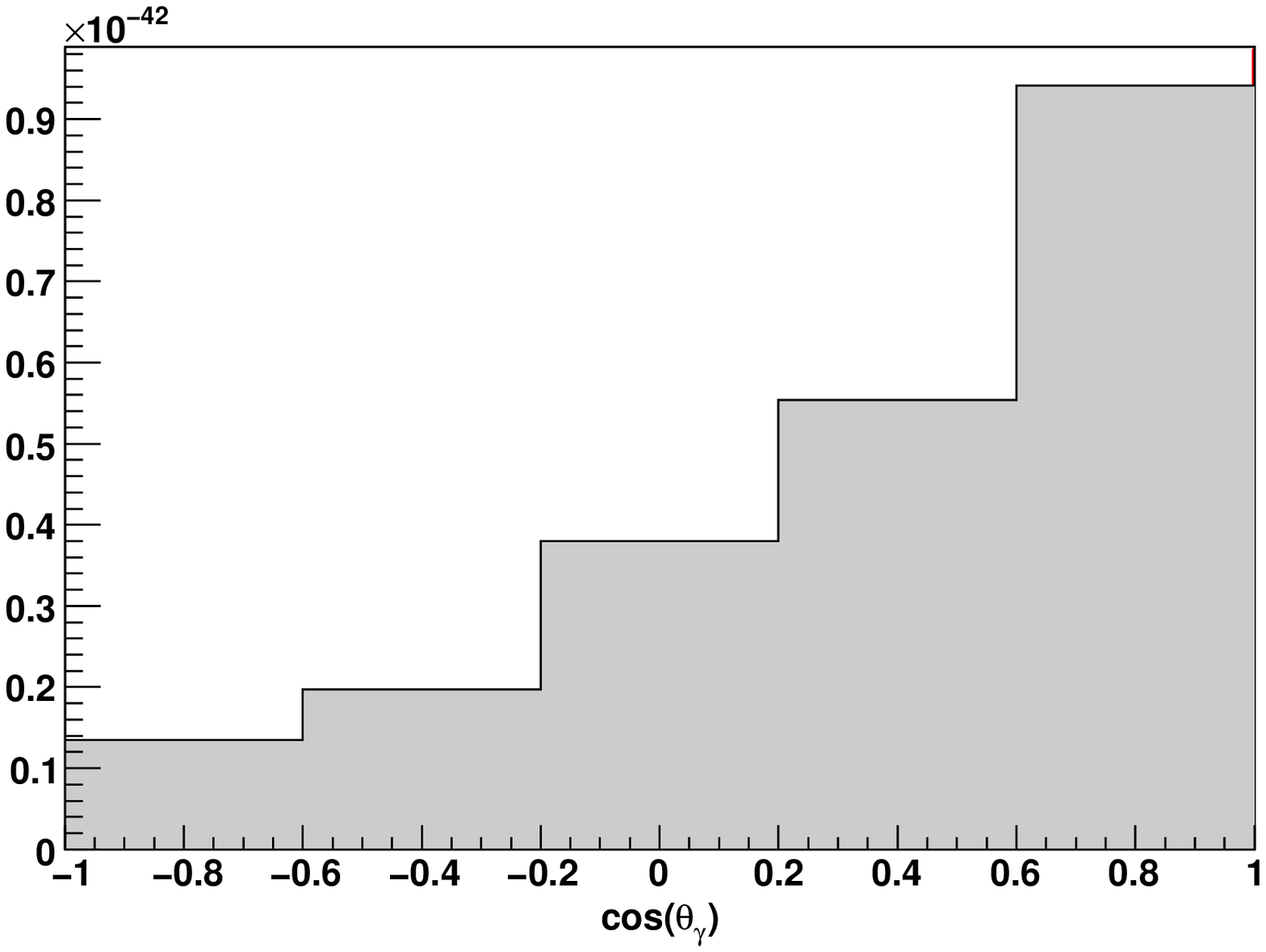}
\includegraphics[width=12pc, height=12pc]{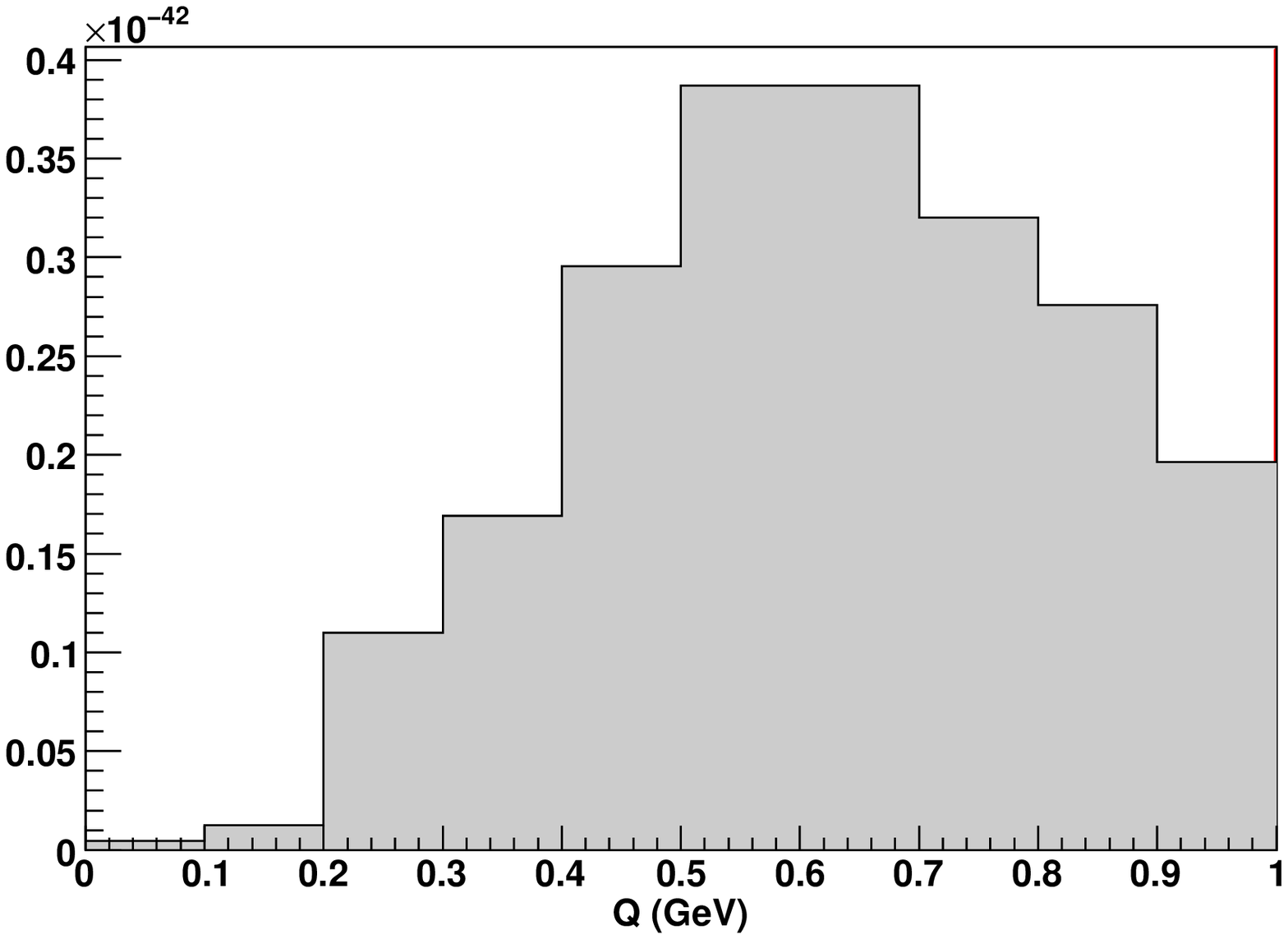}
\caption{
\label{fig:incoh_distrib}
Distributions for photon energy $E_\gamma$, photon angle $\cos\theta_\gamma$, 
and nuclear recoil $Q$, for each of the $\omega$ (top row), $\Delta$ (second row) and Compton-like
(third row) contributions to the incoherent process $\nu p \to \nu p \gamma$ at $E_\nu = 1\,{\rm GeV}$. 
}
\end{center}
\end{figure}
With all of the ingredients in place, it is straightforward to compute
cross sections. 
This section provides a discussion of single-nucleon interactions, 
and the following section turns to coherent interactions
involving compound nuclei.  

Fig.~\ref{fig:full} displays the various contributions to $\nu N \to \nu N\gamma$ 
including the effects of recoil of the final state nucleon, and the form factors 
specified in the previous section.  
The total Compton-like cross section for protons is divergent due to
bremsstrahlung emission of soft photons. For illustration, 
the figure shows the total Compton-like cross section on protons 
for photon energy above $200\,{\rm MeV}$. 
Since suppressed 
isovector contributions have been ignored 
in the meson exchange 
case, and since $\Delta$ resonance production results in pure isoscalar interactions,
the remaining cross sections are the same for protons and neutrons.   

At low energy, the cross sections reduce to the zero-recoil 
expressions derived earlier in (\ref{comptonxs}) and (\ref{zero_omega})%
\footnote{ 
Recall that the $\omega$ and $\Delta$ contributions are of precisely 
the same form in this limit - we have neglected interference terms that would
be important at very low energy.  
}.
In the zero-recoil limit there is no interference 
between vector and axial contributions in the Compton-like case,
so that the cross sections for $\nu$ and $\bar{\nu}$ are identical in this limit.    
The operator describing $\omega$ and $\Delta$ channels
at low energy involves only the axial-vector component of the weak hadronic
current.  
The cross sections for these contributions are therefore 
also identical for $\nu$ and $\bar{\nu}$ in the zero-recoil limit.   
At large energy, interference between vector and axial-vector contributions
yields a larger cross section for neutrinos over antineutrinos, except 
for the meson-exchange case where suppressed weak-vector contributions have been neglected.  

Figure~\ref{fig:incoh_distrib} displays partial cross sections as a function of
photon energy, photon angle, and nuclear recoil $Q=[-(k-k')^2]^{1/2}$, for each
of the $\omega$-induced, $\Delta$-induced and Compton-like cross sections.    These
representative results are for $E_\nu = 1\,{\rm GeV}$ neutrinos scattering on protons.  

\begin{figure}
\begin{center}
\includegraphics[width=20pc, height=20pc]{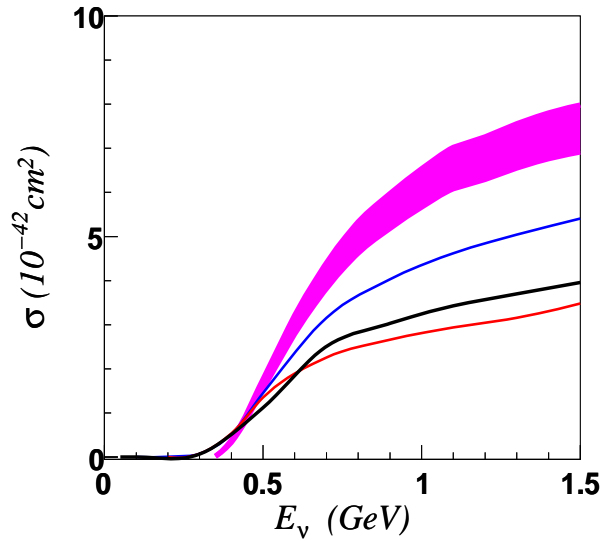}
\caption{
\label{fig:err}
Variation of the single nucleon cross section induced by $\Delta$ resonance
production.  The middle(black) line 
is the same as depicted in Fig.~\ref{fig:full}, 
with $z=0$, energy-independent width and magnetic-dominance form factors. 
The upper (blue) line is for form factor values in (\ref{newff}); 
the lower (red) line is for energy-dependent width (\ref{width}); 
and the (magenta) band is evaluated by multiplying the
total cross section for $\nu N \to \nu \Delta$ by the branching
fraction $0.52 - 0.60 \times 10^{-2}$ for $\Delta\to N\gamma$. 
}
\end{center}
\end{figure}

As an indication of uncertainties for the $\Delta$ contribution,
Fig.~{\ref{fig:err}} shows the cross section calculated with energy-dependent width 
(\ref{width}); and
using an alternate fit for the coefficients in (\ref{dcoeffs}):~\cite{Lalakulich:2006sw}  
\be\label{newff}
C^A_5 = 1.2\,, \quad 
C^V_3 = 2.13 \,, \quad
C^V_4 = -1.51 \,, \quad
C^V_5 = 0.48 \,. 
\ee 
The cross sections for offshell parameter
$z=\pm 1$ in (\ref{eq:z})
are different by more than a 
factor of 2 at $E_\nu\to 0$, as determined by (\ref{eq:zdelta}).  
However, with the default form factor model and in the energy range considered, 
the total cross sections for 
$z=\pm 1$ differ by only a few percent above a few hundred MeV, 
where $\Delta$ can be produced onshell.  
The energy dependence of the width also has relatively minor impact on the total cross section
above a few hundred MeV, 
suggesting that offshell effects do not impact GeV-scale cross sections dramatically.
The spread in these curves in Fig.~\ref{fig:err} can be taken as a crude estimate of the 
cross section uncertainty.   
Also displayed in Fig.~\ref{fig:err} is the result obtained by multiplying
the total cross section for $\nu N \to \nu \Delta$ (either $\nu p \to \nu \Delta^+$ or $\nu n \to \nu \Delta^0$) 
by the branching fraction $\sim 0.52-0.60$~\cite{Amsler:2008zzb} 
for $\Delta \to N\gamma$.   
For this case, the $\Delta$ width 
is ignored, and the default form factors (\ref{dcoeffs}) are used%
\footnote{
The values for the total width and photon 
branching fraction obtained from form factors in (\ref{dcoeffs}) 
are $120\,{\rm MeV}$ and 0.42. For 
the form factors (\ref{newff}) the total width is unchanged 
and the photon branching fraction becomes 0.52. 
}.   

Corrections to the incoherent single-nucleon cross-sections from 
nuclear effects such as Fermi motion and Pauli blocking have been neglected. 
These considerations are not unique to single-photon production cross sections, 
and are beyond the scope of this paper.  These effects 
should be incorporated in a more precise analysis, but are not expected to be 
dramatic for relatively large neutrino energies ($E\sim 1\,{\rm GeV}$) 
on relatively small nuclei (e.g. $^{12}$C).   
The possibility of coherent processes, where the nucleus stays intact, 
is a distinct and interesting possibility.  This is the subject of the
following section. 

\section{Coherence effects \label{sec:coh}} 

In addition to interactions with individual nucleons, the weak and electromagnetic 
currents can scatter coherently off an entire nucleus.  
This section investigates the coherent contributions for each of the Compton-like, 
$\omega$-induced and $\Delta$-induced processes.  

\subsection{Compton scattering} 

\begin{figure}
\begin{center}
\includegraphics[width=10pc, height=10pc]{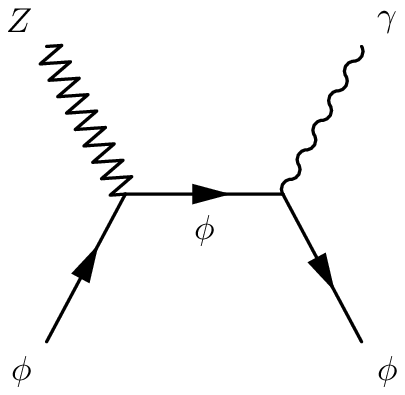}
\hspace{10mm}
\includegraphics[width=10pc, height=10pc]{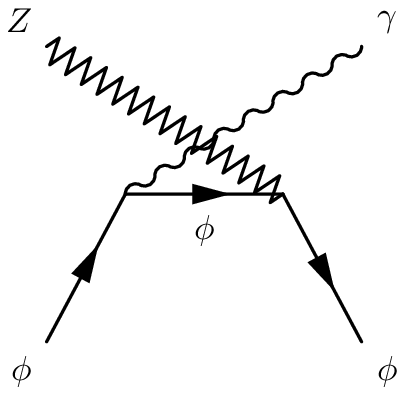}
\hspace{10mm}
\includegraphics[width=10pc, height=10pc]{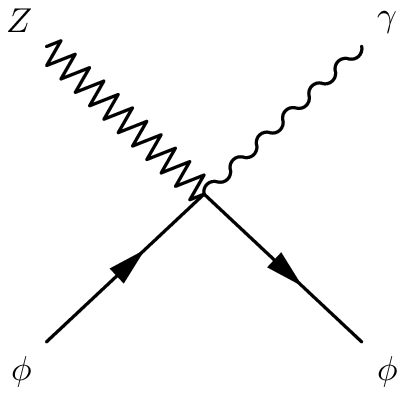}
\caption{\label{fig:seagull}
Coherent component of generalized Compton scattering off
a compound nucleus.
}
\end{center}
\end{figure}

The vector coupling of the $Z$ boson is primarily to the neutron, 
due to the smallness of the factor $1-4\sin^2\theta_W \approx 0.08$
appearing in the proton coupling.
Of course, the vector photon couples directly only
to the charged proton.    Thus, the Compton-like cross section on 
either an isolated proton, or an isolated neutron, is smaller than 
naive power counting suggests. 
However, when coherent effects are considered, the $Z$ and photon 
couple to the {\it total} vector weak charge, and the {\it total} 
electric charge, respectively, and the charge suppressions are 
no longer effective. 
This process can be viewed as initial- and final-state radiation 
from the as-yet unobserved coherent neutral-current scattering of 
a neutrino from an intact, recoiling nucleus~\cite{Freedman:1973yd}.   

The scattering process can be described simply in the case of a
spinless, isoscalar nucleus, e.g. ${\cal N} = ^{12}$C.  
Only isoscalar couplings are relevant, 
and the interactions with vector $Z$ and photon are described at low energy 
by an effective scalar field,
with Lagrangian 
\be\label{Lphi}
{\cal L} = |D_\mu \phi|^2 \,,
\ee
where the covariant derivative is 
\be
D_\mu \phi = \partial_\mu \phi - i \left( 
e Q A^{\rm e.m.}_\mu + {g_2\over 2\cos\theta_W} Q_W Z_\mu \right) \phi \,. 
\ee
The electric charge for an isoscalar nucleus is
\be 
Q = Z = \frac12 A \,,
\ee
and the weak vector charge is ($Z=N=A/2$)
\be
Q_W = Z ( \frac12 - 2s_W^2) + N ( -\frac12 ) = - s_W^2 A  \,.
\ee
For the nonradiative process $\nu {\cal N} \to \nu {\cal N}$, the cross section 
at low energy is calculated from (\ref{Lphi}) to be~\cite{Freedman:1973yd}
\be
\sigma = {1\over \pi} G_F^2 A^2 s_W^4 E^2 \,. 
\ee
As the energy is increased,
coherence becomes confined to the region 
of small momentum transfer, as implemented by including a form factor, 
\be
d\sigma \to d\sigma |F( (k-k')^2 )|^2  \,.
\ee
Neglecting asymmetries in the neutron and proton distributions, this 
form factor should be the same as measured in electromagnetic scattering on 
the nucleus, $e^- {\cal N} \to e^- {\cal N}$.  The phenomenological 
form $F(t)=\exp(bt)$ is adopted, 
where for $^{12}$C we take $b\approx 25\,{\rm GeV}^{-2}$~\cite{Freedman:1973yd}. 
In general nuclei, $b$ is expected to scale as $b \sim \langle r^2 \rangle \sim A^{2/3}$.  

A straightforward calculation of the diagrams in Fig.~\ref{fig:seagull} 
shows that the Compton-like cross section at very low energy is
\be\label{eq:cohcom}
{d\sigma\over dedx} = {\alpha G_F^2 E^4 \sin^4\theta_W A^2 \over 4\pi^2 m_N^2 } 
e (1-e) \left[ 
{1\over e^2}\left(\frac12 -\frac16 x^2\right) 
+ \frac{1}{e} \left( -\frac76 + \frac56 x^2\right)  + \frac43 - \frac23 x^2 - \frac23 e \right] \,. 
\ee
In fact, this can be recognized as the cross section for a 
fermion of electric charge $F^1=Q$, weak charge $C_V=Q_W$ and 
mass $A m_N$, cf. (\ref{comptonxs}).  This should be true, since the low-energy 
electroweak probes cannot tell whether a single nucleon 
carries both $Q$ and $Q_W$, or whether the charge is distributed over
different nucleons.   
As the energy is increased, the coherent amplitude is again 
restricted to small momentum transfers.   
This is implemented by introducing the same form factor as above,
so that the cross section is obtained by replacing the 
single-nucleon cross section by the ansatz
\be\label{ff}
d\sigma(A) \approx A^2  e^{2b(k-k')^2} d\sigma(1)\,.
\ee

\subsection{Virtual meson exchange}

We have noted that $c^{(3)}_4$ in (\ref{eq:expansion}) 
describes the leading operator providing nuclear coherence for the axial-vector weak current.  
Neglecting nuclear modifications to the meson couplings, 
the coherent cross section induced by $\omega$ exchange 
at very low energy is simply $A^2$ times
the single nucleon cross section (\ref{zero_omega}).
We again adopt a gaussian form factor, reflecting the distribution 
of nucleons inside the nucleus, and the cross section is
modified from the single-nucleon case according to (\ref{ff}). 

\subsection{Coherent resonant production}

\begin{figure}
\begin{center}
\includegraphics[width=20pc, height=20pc]{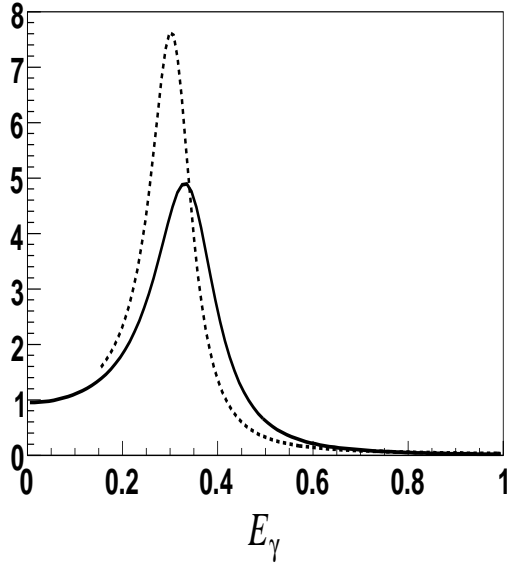}
\caption{
\label{fig:factor_coh}
Resonant enhancement factor for coherent scattering induced
by $\Delta$ resonance production.  The solid line is for a 
fixed width $\Gamma_\Delta=120\,{\rm MeV}$, the dashed line for
an energy-dependent width, cf. (\ref{width}).  
}
\end{center}
\end{figure}

When the photon energy is such that 
\be
m_\Delta^2 - (k'+q)^2 \approx m_\Delta^2 - m_N^2 - 2m_N E_\gamma \approx 0 \,,
\ee
a resonant effect comes into play.   Taking into account the 
finite width, the effect can be described by introducting a factor, 
for nonrelativistic nucleons, 
\begin{align}\label{resenhance}
{1\over (m_\Delta^2-m_N^2)^2} &\to
\left| \frac12 
 {1\over m_\Delta^2 - m_N^2 - 2m_N E_\gamma - i\Gamma_\Delta m_\Delta } 
+ 
\frac12
 {1\over m_\Delta^2 - m_N^2 + 2m_N E_\gamma - i\Gamma_\Delta m_\Delta } 
\right|^2
\nl
&= {1\over (m_\Delta^2-m_N^2)^2} \times 
{ 1 + \gamma^2 \over (1-\beta^2 - \gamma^2)^2 + 4\gamma^2} \,,
\end{align}
where $\gamma=m_\Delta \Gamma_\Delta/(m_\Delta^2-m_N^2)$ and 
$\beta = 2m_N E_\gamma/(m_\Delta^2 -m_N^2)$.  
The enhancement factor is plotted in Fig.~\ref{fig:factor_coh}.
Coherence is again implemented by the ansatz (\ref{ff}). 

The coherent aspect of the $s$-channel $\Delta$ contribution 
can be understood as follows. 
In the limit $m_\Delta , m_N \gg m_\Delta - m_N$, the matrix
elements can be calculated for nucleons at rest, and  
the leading interactions of the $N-\Delta$ system 
with vector and the axial-vector currents 
are proportional to the isovector magnetic moment operator.   
This can be seen explicitly by taking the nonrelativistic limit of 
the equation of motion derived from (\ref{eq:delta_lagr}), leading to 
\be
i \epsilon^{ijk} \gamma^j \gamma_5 \Delta^{a,k} \sim \Delta^{a,i} \,. 
\ee
Thus for neutral fields, the hadronic part of the operators for 
$c^{(1)}_{N\Delta}$ and $c^{(2)}_{N\Delta,1}$
in (\ref{eq:DNLO}) and (\ref{deltaff}) both involve $\bar{\Delta}^{a=3}_i N$, 
plus terms suppressed by powers of $1/m_N$.     
For each component of the external field (e.g. $A^{a,i}$ in (\ref{eq:DNLO}), or 
$\epsilon^{ijk} F^{a, jk}$ in (\ref{deltaff}) ) 
the excitation $N\to \Delta$ occurs with fixed amplitude 
between a nucleon in a given 
spin state 
and a unique corresponding $\Delta$ spin state. 
The de-excitation $\Delta\to N$ obeys the same selection rule.
For excitation by the weak neutral current, and de-excitation 
through the magnetic field, 
this leads to a $\bm{J}\cdot \bm{B}$ interaction coupled coherently 
to the nucleon.  
If the scattering takes place on a collection of nucleons, 
the final state cannot distinguish which nucleon was struck,
giving rise to coherence.  

An insightful model of the static nucleon transitions is obtained  
by viewing the baryons as solitonic ``twisted'' configurations of the pion field 
carrying unit baryon number. 
The solitonic description for large (odd) $N_c$ 
predicts a multiplet of low-lying baryons with 
spin and isospin $I=J=1/2, 3/2, \dots , N_c/2$.  In particular for $N_c=3$, 
the nucleon and $\Delta$ are singled out.   
The couplings of the baryons to external fields take a unique form, 
specified by the operators,  
\be\label{skop}
O^{i,a} = 
{\rm Tr}( \tau^i a^{-1} \tau^a a ) \,. 
\ee
Here $a = a^0 + \bm{a}\cdot \tau$ is an $SU(2)$ matrix field, and the baryon 
lagrangian is defined on the space of coordinates $\{a^0,a^1,a^2,a^3\}$.  
An explicit realization for the ``potential'' on this space of coordinates 
is the Skyrme model~\cite{Skyrme:1961vq,Adkins:1983ya}, although many 
relations, such as properties following from the uniqueness of (\ref{skop}), 
are model-independent predictions of the large $N_c$ limit. 

It is interesting to consider the coherent enhancement 
in terms of a simplified two-state model of the nucleon-$\Delta$ system. 
In this context, the excitation takes the ground state to a coherent
superposition of singly-excited states for each of the $A$ nucleons, 
\be
|\downarrow \downarrow \downarrow \dots \rangle \to 
{1\over \sqrt{A}}\left( 
|\uparrow \downarrow\downarrow \dots \rangle 
+ |\downarrow \uparrow \downarrow \dots \rangle + \dots 
\right)\,,
\ee
with amplitude proportional to $\sqrt{A}$.  
Similarly, the amplitude for de-excitation and emission of a 
photon from this state is 
proportional to $\sqrt{A}$.  Hence the amplitude for the 
total process grows as $A$, and the cross section as $A^2$.
This resonant coherent process has some relation to 
the ``Dicke superradiance'' effect encountered 
in atomic physics~\cite{Dicke:1954zz}%
\footnote{
I thank B. Adams for pointing out this reference. 
}.
However, in the language of spin systems, our spin $j=A/2$ system is
prepared in $|j=A/2, j_z=-A/2+1\rangle$. 
The true superradiant effect occurs when the system is somehow 
prepared in $|j=A/2, j_z\approx 0\rangle$, due to the large 
coupling $\langle j j_z-1| \sigma_- | j j_z\rangle = \sqrt{(j+j_z)(j-j_z+1)} \sim j \sim A$ 
for $j_z\sim 0$ (versus $\sqrt{A}$ for emission from the singly excited state).  
It would be amusing to consider whether such ``nuclear superradiance'' 
could be observed in practice, perhaps in some extreme astrophysical 
environment (it is certainly inefficient to induce multiple excitations by weak currents).
It is also interesting to investigate the impact of density and temperature dependence of 
the $N-\Delta$ mass splitting. 

\subsection{Breakdown of coherence}

Coherence is restricted to the case of momentum transfers 
that are small compared to the inverse size of the nucleus, 
with linear dimension $\sim A^{1/3}$.  
For moderately sized nuclei, the naive $A^2$ scaling of 
the zero-recoil cross-sections is significantly  modified  
already at hundreds of MeV incident neutrino energies. 
Before plotting the final cross sections, it is instructive to examine in some detail 
the limit of large energy, or large nucleus (specifically large $A^{2/3} E^2$)
to see what remains of the coherent cross section.  

To begin, we notice that the argument of the exponential factor in (\ref{ff}) 
may be expanded as 
\begin{align}\label{arg}
&2b(k-k')^2 = 2b(p-p'-q)^2
= -4bE^2\left[ (1-e)(1-y )
+e(1-x)
-e(1-e)(1-z)
\right] \nl
&\quad = 
-4bE^2 \bigg\{ (1-e)(\sqrt{1-x^2}\sqrt{1-z^2}+xz-y)+\frac12[ (1-e)\sqrt{1-z^2}-\sqrt{1-x^2} ]^2
\nl
&\qquad 
+ \frac12 [ (1-e)(1-z) - (1-x) ]^2 
\bigg\} \,,
\end{align}
where 
$e=E_\gamma/E$, 
$x=\cos\theta_\gamma=\hat{\bm p}\cdot\hat{\bm q}$, 
$y=\hat{\bm p}\cdot\hat{\bm p}'$, 
$z=\hat{\bm p}'\cdot\hat{\bm q}$.
This factor will be small when $\bm{p}'+\bm{q} \approx \bm{p}$, where the equality 
$|\bm{p}'|+|\bm{q}|\approx |\bm{p}|$ is already enforced by the small-recoil limit.  
For this to happen, either  $|\bm{q}|/|\bm{p}|=e$ or $|\bm{p}'|/|\bm{p}| = 1-e$ must be small, 
or the vectors must all be collinear.   
The overall size of the cross section then depends on the behavior 
of the remaining amplitude in these restricted regions of phase space.  

The two cases of interest will be when $e\approx 0$, or when both $e$ and $(1-e)$ are order unity. 
In the first case (soft photon), (\ref{arg}) reduces to 
\be\label{soft}
2b(k-k')^2 \approx -4bE^2(1-y) \approx -2bE^2 \theta_{pp'}^2 \,. 
\ee
which restricts the phase space to $\theta_{pp'} \lesssim (bE^2)^{-1/2}$.   In the 
remaining matrix element we can set $x\approx z$ and $y\approx 1$. 

In the second case (collinear photon), it is convenient to introduce $\theta'$, $\phi'$ 
as polar and azimuthal angles of $\bm{p}'$ with respect to $\bm{q}$, where 
$\bm{p}$ and $\bm{q}$ define the x-z plane.   Then $\sqrt{1-x^2}\sqrt{1-z^2}+xz-y 
= \sqrt{1-x^2}\sqrt{1-z^2}(1-\cos\phi')$, 
$\sqrt{1-x^2}=\sin\theta_\gamma$ and $\sqrt{1-z^2}=\sin\theta'$.  Requiring that each of
the three positive definite terms in (\ref{arg}) is not larger than order unity, the 
relevant region of phase space is 
\be\label{collinear}
\phi' \lesssim {(bE^2)^{-1/2}\over (1-e)\theta'} \,,\quad
\theta - (1-e)\theta' \lesssim (bE^2)^{-1/2} \,,\quad
\theta' \lesssim {(bE^2)^{-1/4} \over \sqrt{e(1-e)} } \,.
\ee
In the remaining matrix element we can set $x\approx y\approx z \approx 1$.   
Retaining the $e$ dependence in (\ref{collinear}) is not essential, but 
will allow us to indicate the leading behavior at $e\to 0$ and $e\to 1$ 
when e.g. $bE^2 \ll e \to 0$.  

\subsubsection{Compton-like process}

Consider first the case of Compton-like scattering.  
The cross section is
\begin{align}
{d\sigma({\rm Compton}) \over de dx} 
&\propto 
A^2 E^4 e(1-e) 
\int d\cos\theta' 
\int d\phi'  
e^{ 2bE^2(k-k')^2  } \bar{\Sigma}|{\cal M}|^2 
\,,
\end{align}
where the spin-averaged matrix element is
\begin{multline}
\bar{\Sigma}|{\cal M}|^2 \propto 
{1-e\over e^2} \big\{ 
-2(1-e)y^2 + y[ -(1-e)^2z^2 + 2(1-e)xz -x^2-e^2+4(1-e) ] \\
-(1-e^2)z^2 +2(1-e)xz -(1-2e)x^2 +2(1-e) + e^2 \big\} \,. 
\end{multline}
At small $bE^2$, the gaussian factor can be neglected, and the integral 
over the final-state neutrino angle yields the result (\ref{eq:cohcom}).  

If we look for contributions at large $bE^2$ 
with $e$ and $1-e$ order unity, i.e., where
both the photon and final-state neutrino carry a substantial fraction of
the incoming neutrino energy, the relevant region of phase space is (\ref{collinear}).  
Taking the appropriate limit of the matrix element, 
\be
\sigma({\rm Compton}) \sim A^2 E^4 (bE^2)^{-3/2} \int de {(1-e)^2\over e^2} \sim A E \int de {(1-e)^2\over e^2}\,.
\ee
For example, the partial cross section for photon carrying more than a fixed fraction 
of the incident neutrino energy scales as $A E$, with the photon emitted in the forward 
direction. 

In contrast, the partial cross section 
for photons in a fixed energy range corresponds to the case $e = E_\gamma/E \to 0$.    
For this case, (\ref{soft}) applies, and the cross section becomes
\be
{d\sigma({\rm Compton}) \over dx} 
\sim A^{4/3} E^2 \int {de\over e} \,.
\ee
The cross section for small-energy photons is dominant for large $bE^2$, scaling as $A^{4/3}E^2$. 
In this limit, 
the cross section is flat in photon angle, and logarithmically divergent for arbitrarily small
photon energy.  

\subsubsection{$\omega$ and $\Delta$ processes}

Moving next to the interaction induced by $\omega$ or $\Delta$, 
the cross section takes the form (apart from a possible coherent enhancement depending
on $e$)
\begin{align}
{d\sigma(\omega\,{\rm or}\,\Delta) \over de dx} 
&\propto 
A^2 E^6 e(1-e) \int d\cos\theta' \int d\phi' 
e^{ 2b(k-k')^2}  \bar{\Sigma}|{\cal M}|^2 
\,,
\end{align}
with 
\be\label{omegam2}
\bar{\Sigma}|{\cal M}|^2 
\propto
e^2(1-e) ( 1- xz ) \,. 
\ee
Again, at small energy the gaussian factor can be ignored, and 
the integral over final-state neutrino angle yields a cross section 
of the form (\ref{zero_omega}).  

When the fractional energy, $e$, carried by the photon goes to zero, (\ref{soft}) applies, 
and the cross section takes the form
\be\label{deltcoh}
{d\sigma(\Delta) \over de dx} \sim A^2 E^6 e^3 (1-x^2) \int dy e^{-4bE^2(1-y)}
\sim 
A^{4/3} E^4 e^3 (1-x^2) \,.
\ee
For example, for the $\Delta$ contribution the photon energy is tied
to the resonance mass, $e\sim (m_\Delta^2-m_N^2)/2m_NE \to 0$.  
Integrating over $e$, the 
total cross section becomes independent of the incoming neutrino energy, and 
develops a $1-x^2$ angular dependence: $d\sigma/dx \sim A^{4/3} (1-x^2)$.  

When both $e$ and $1-e$ are order unity (\ref{collinear}) applies, and 
the cross section scales as 
\be\label{omegacoh}
{d\sigma(\omega)\over de} 
\sim A^{2/3} E^2 e [ 1 + (1-e)^2 ]\,.
\ee
The photon is emitted in the forward direction, with energy spectrum 
tilted towards $e=1$.   
At large nuclear size, the cross section grows as $A^{2/3}$.   
This case is relevant to the $\omega$ exchange process, where the photon 
is able to carry an arbitrary fraction of the incoming neutrino energy.    

Note that form factors in the 
vector or axial-vector channel of the weak current 
will induce an additional suppression involving powers of 
\be
[ 1 - (p-p')^2/m_{V,A}^2 ]^{-1} =  [ 1 + E^2(1-e)(1-y)/m_{V,A}^2 ]^{-1}  \,.
\ee
In regions where $y\to 1$ (final-state neutrino collinear), 
this suppression is postponed to a higher scale.  For example, 
for the $\omega$ process, we have $1-y \sim (bE^2)^{-1/2}$, so that
the form factor cuts off energies above a mass scale parametrically 
of order $m_{\rm eff} \sim m_{A}^2 b^{1/2} \sim A^{1/3} m_A^2 /({1\, \rm GeV})$. 
The restricted coherent cross section (\ref{omegacoh}) 
grows more slowly than the number of nucleons $A$, and hence more slowly as a function 
of $A$ than the incoherent process.  However,  
the scale at which $E^2$ growth cuts off also becomes larger with $A$.  

From these considerations, the coherent cross sections are expected 
to behave very differently at large values of $bE^2$.   
The total Compton-like process grows asymptotically as 
$A^{4/3} E^2 \log{E/E_{\rm min}}$ above a threshold photon energy $E_{\rm min}$,
with photon energy spectrum weighted at the low end, and a flat distribution 
in photon angle.      
The $\Delta$ contribution saturates as a function of energy, 
and grows asymptotically with nuclear size as $A^{4/3}$.
The photon energy is fixed by the $\Delta$ excitation energy,
and there is a $1-x^2$ photon angular distribution in the asymptotic limit.  
Finally, for the $\omega$-mediated process, the growth with 
energy and nuclear size is $A^{2/3} E^2$, 
and the process favors a forward photon carrying a large fraction of
the incident neutrino energy.  
The $1\,{\rm GeV}$ energy range for a medium-sized nucleus like 
$^{12}C$ is in a transition region
from small to large $bE^2$, 
but the asymptotic limits are a useful guide for understanding the 
qualitative features of the coherent cross sections. 

\subsection{Summary of coherent single photon cross sections}

\begin{figure}
\begin{center}
\includegraphics[width=20pc, height=20pc]{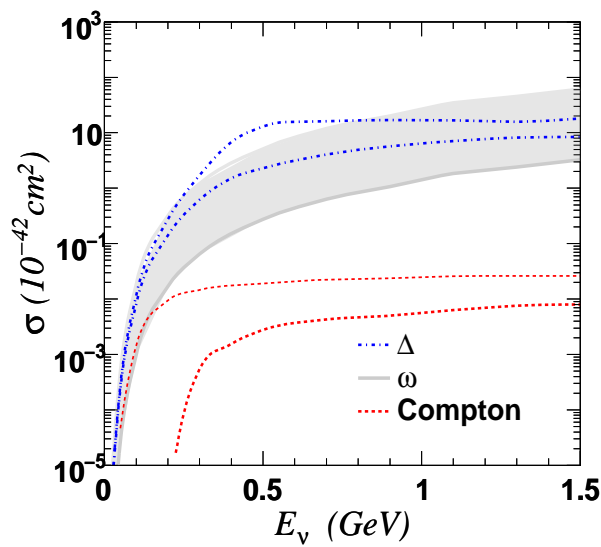}
\caption{
\label{fig:coherent}
Coherent cross sections on $^{12}$C 
induced by Compton-like process; 
$t$-channel $\omega$; 
and $s$-channel $\Delta$.   
A cut $E_\gamma\ge 20\,{\rm MeV}$ (top) and $E_\gamma \ge 200\,{\rm MeV}$ (bottom) 
is placed on the infrared-singular
Compton-like cross section. 
For $\Delta$, separate $\nu$ (top) and $\bar{\nu}$ (bottom) cross sections are shown.   
The band represents the combined effect of $\omega$ and $\Delta$ if resonant structure is
ignored. 
}
\end{center}
\end{figure}

\begin{figure}
\begin{center}
\includegraphics[width=12pc, height=12pc]{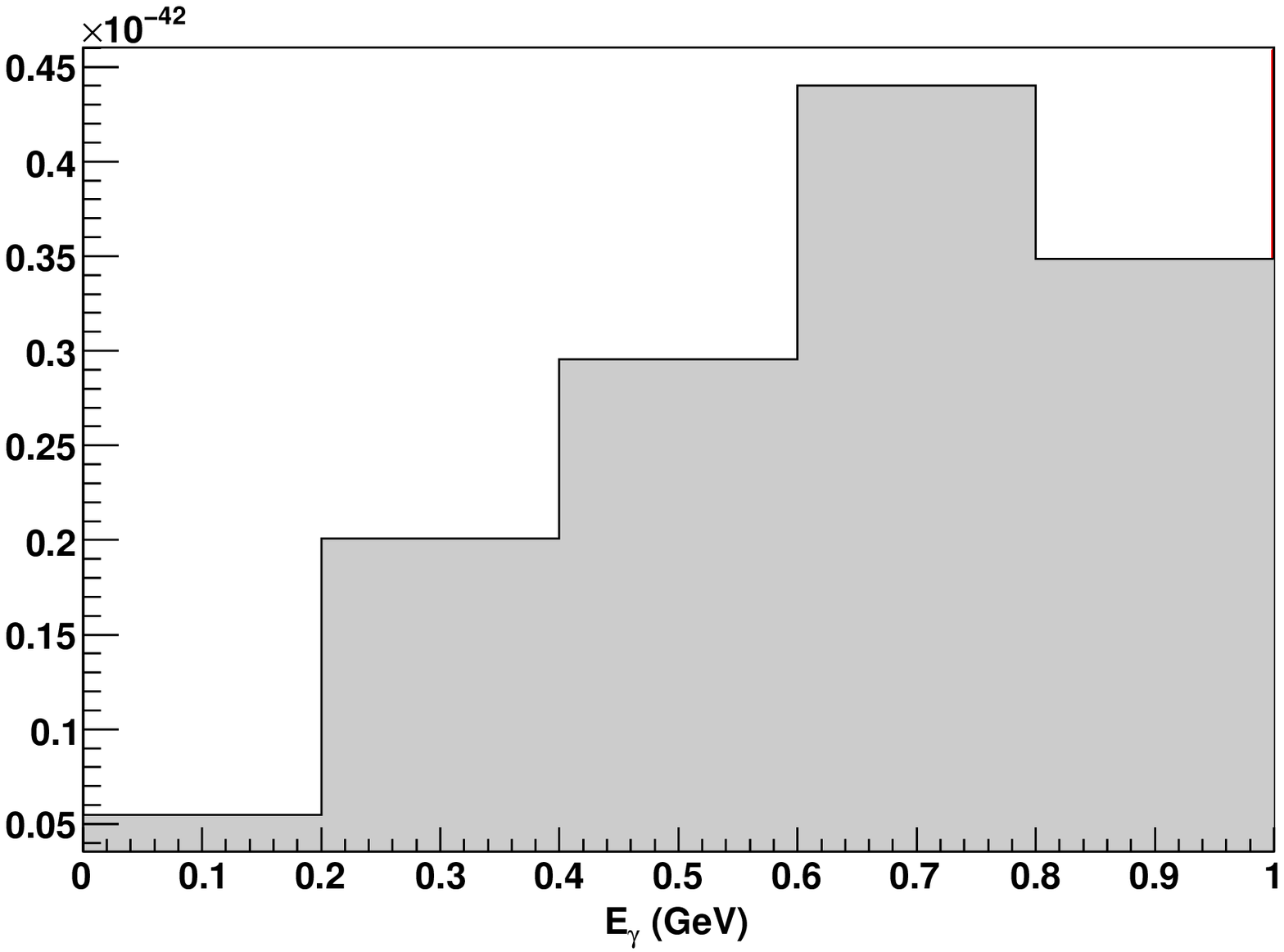}
\includegraphics[width=12pc, height=12pc]{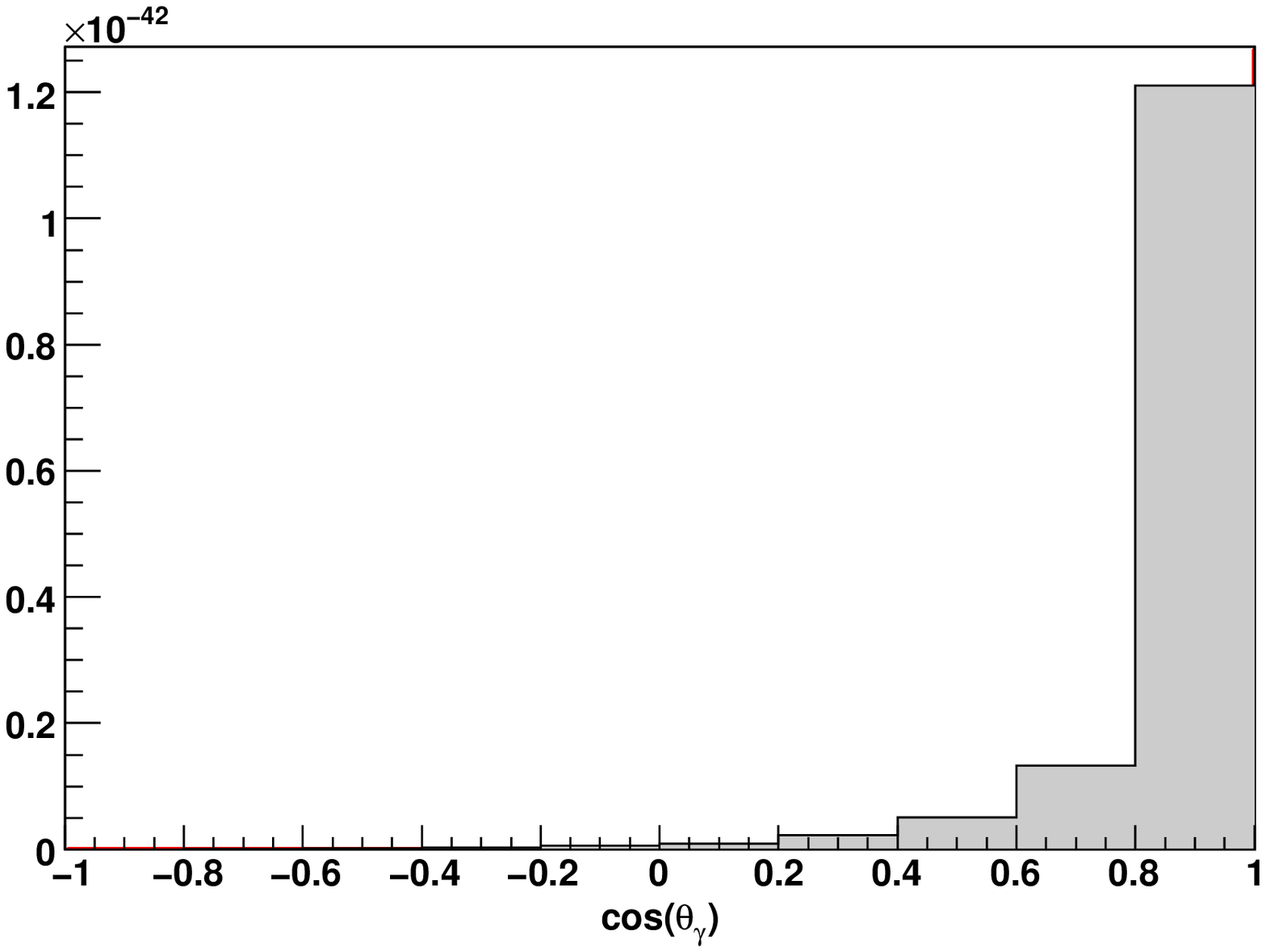}
\includegraphics[width=12pc, height=12pc]{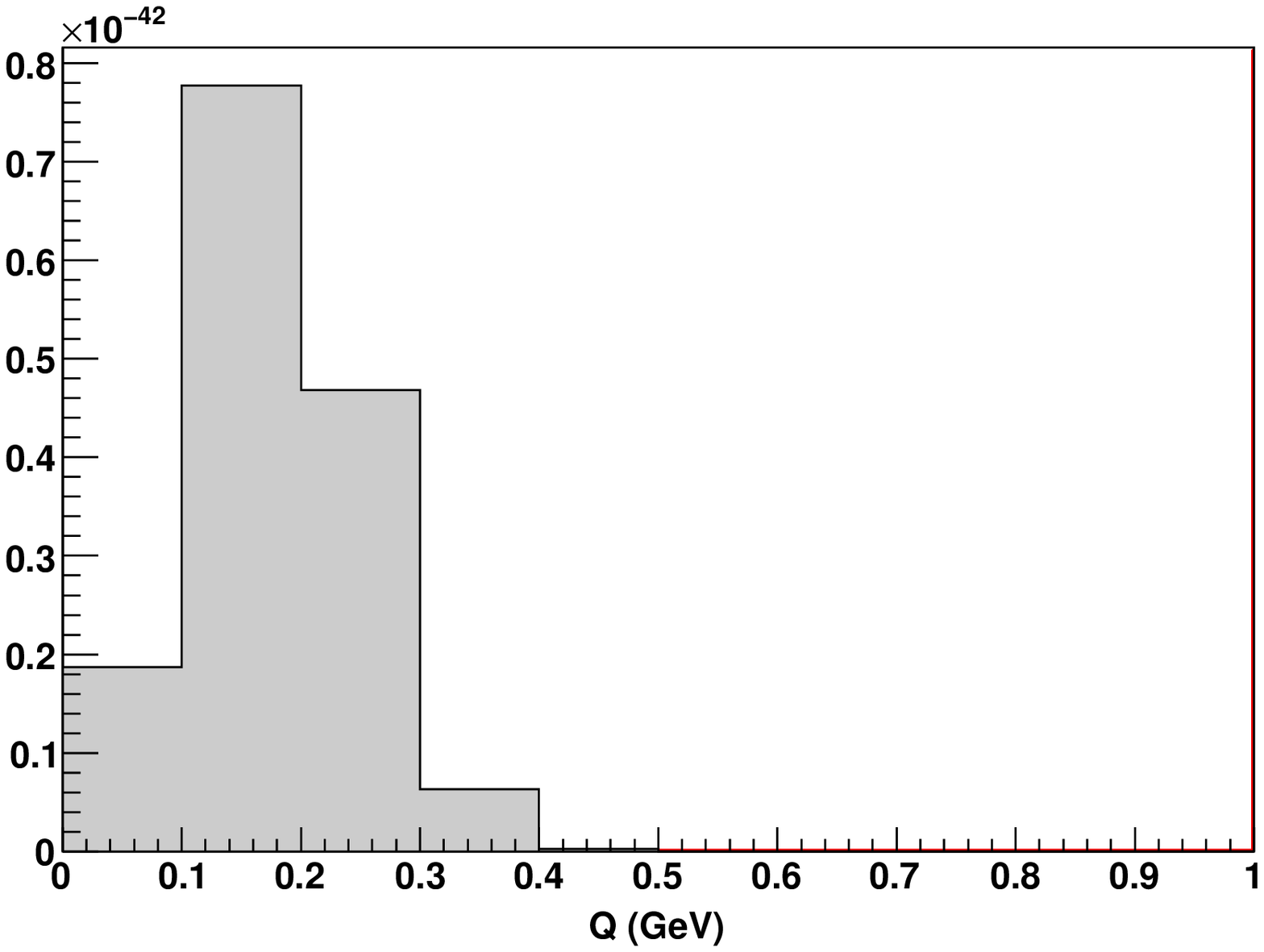}
\includegraphics[width=12pc, height=12pc]{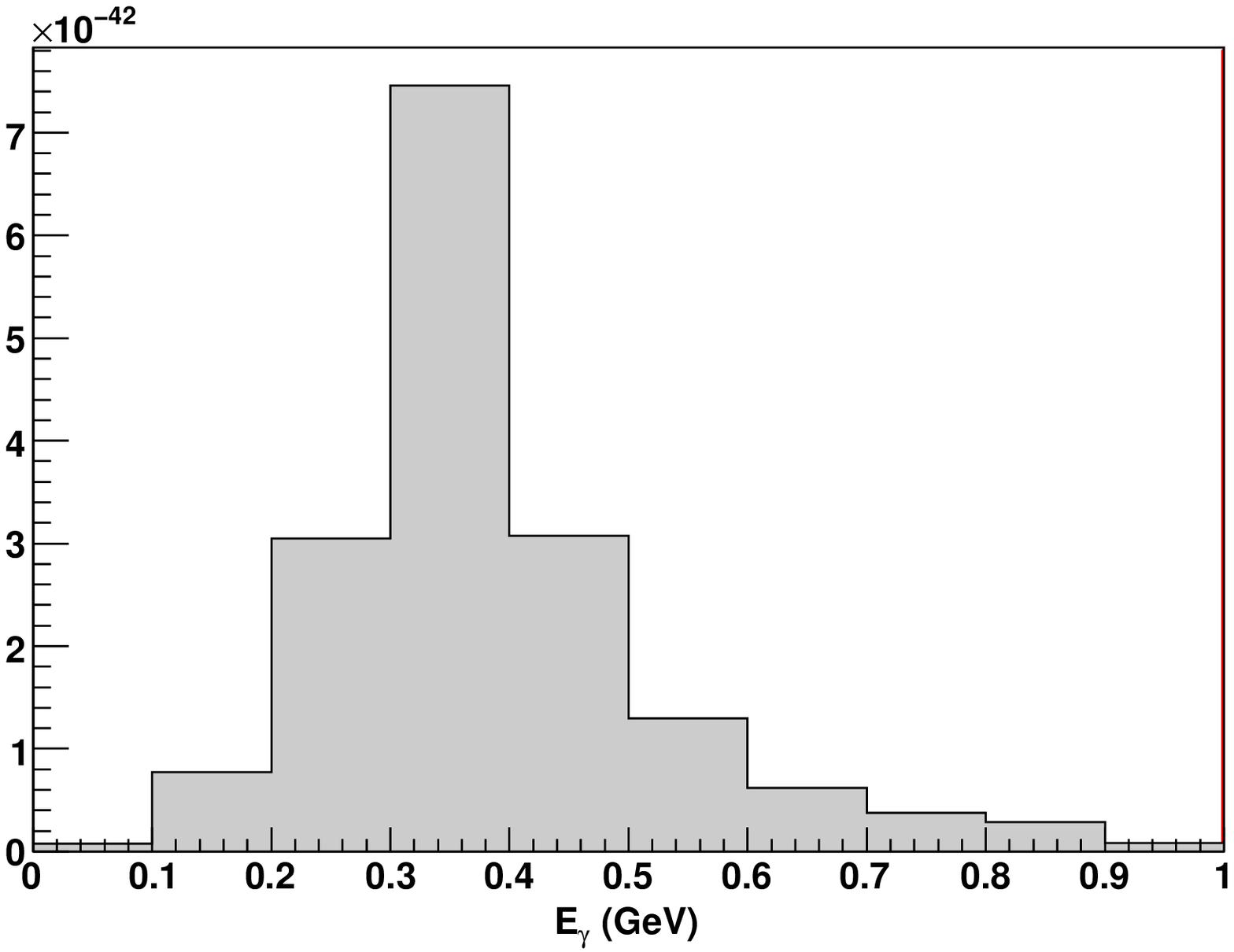}
\includegraphics[width=12pc, height=12pc]{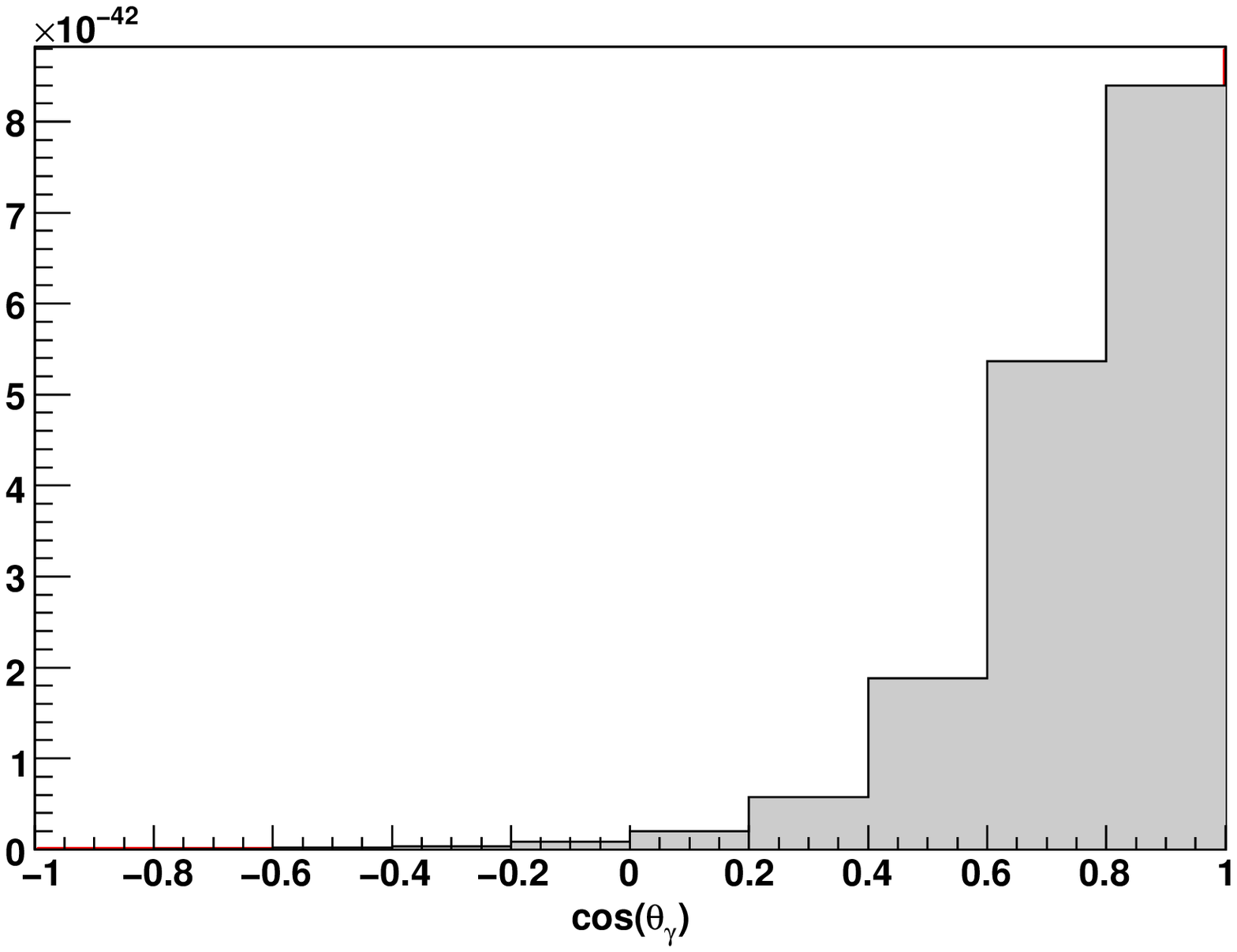}
\includegraphics[width=12pc, height=12pc]{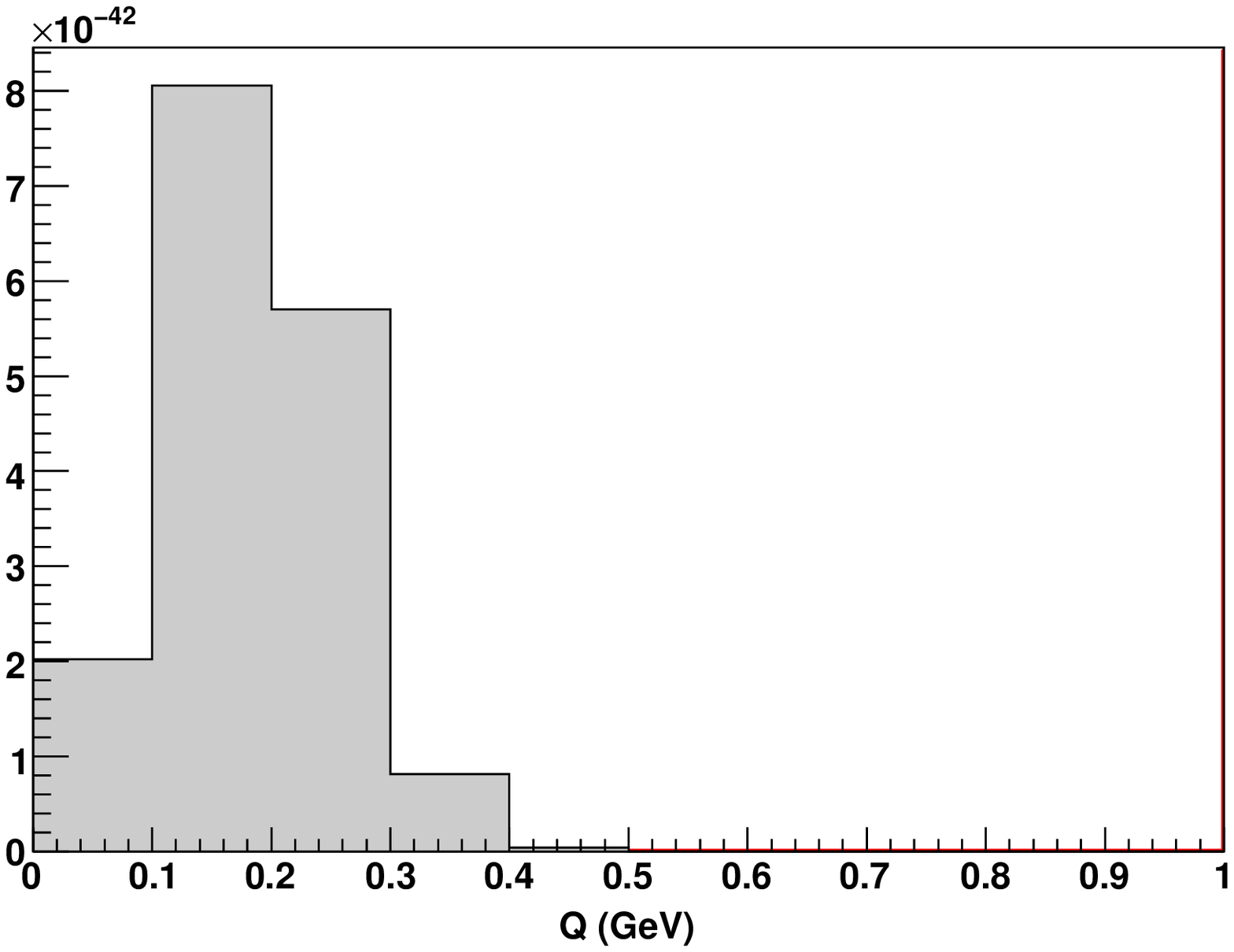}
\includegraphics[width=12pc, height=12pc]{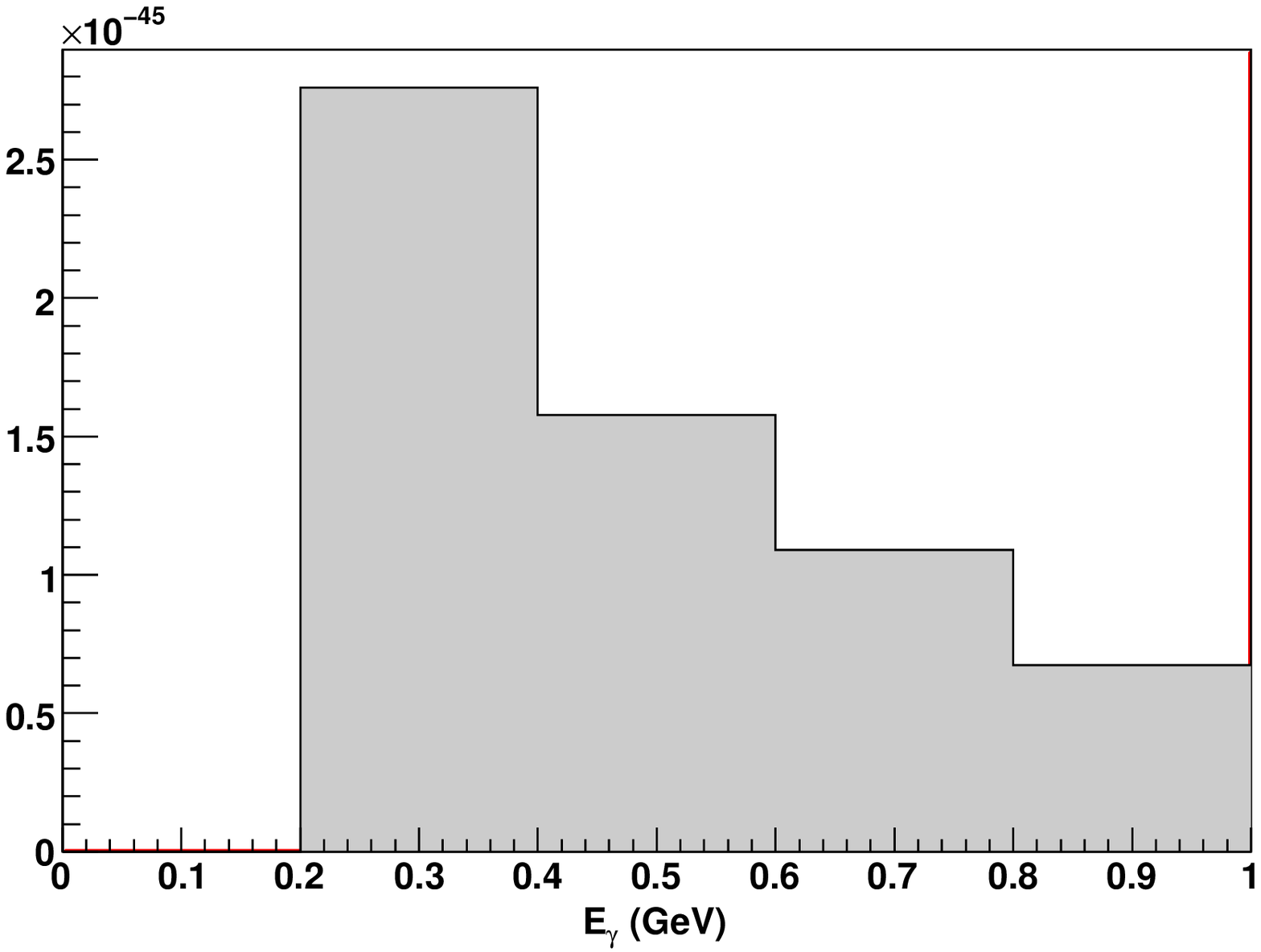}
\includegraphics[width=12pc, height=12pc]{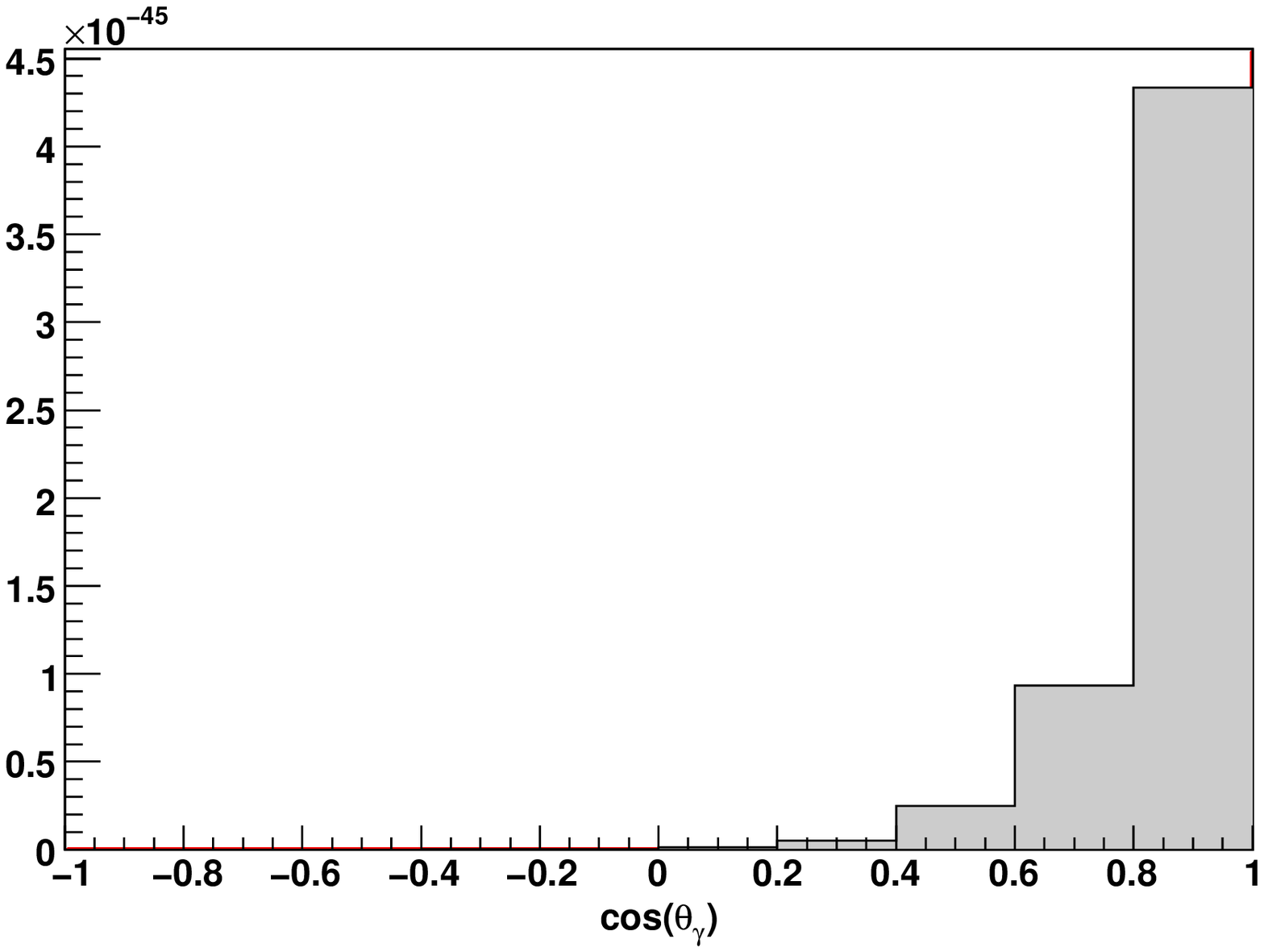}
\includegraphics[width=12pc, height=12pc]{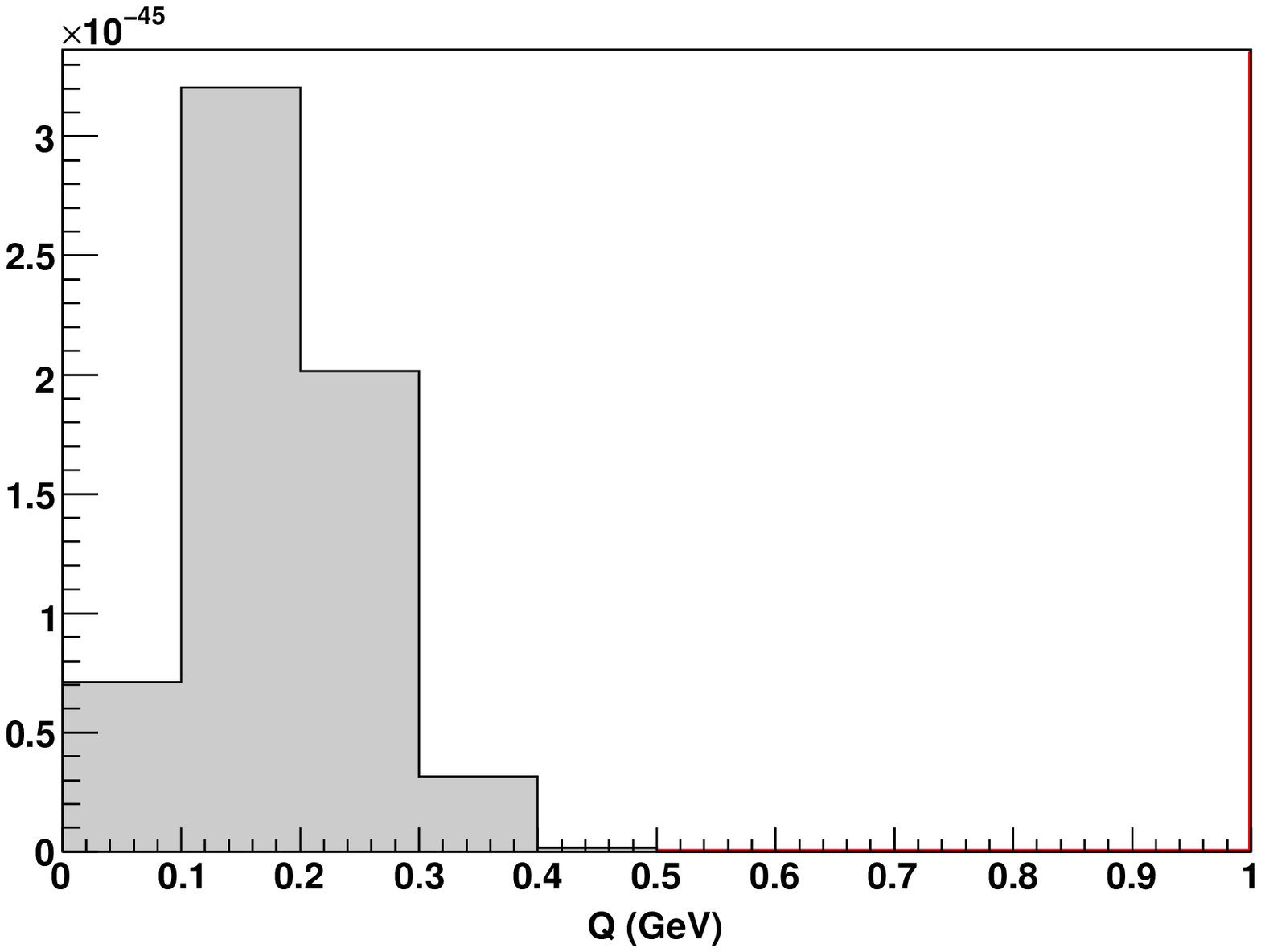}
\caption{
\label{fig:distrib}
Distributions for photon energy $E_\gamma$, photon angle $\cos\theta_\gamma$, 
and nuclear recoil $Q$, for each of the $\omega$ (top row), $\Delta$ (middle row)  
and Compton-like (bottom row)
contributions to coherent process $\nu {\cal N} \to \nu {\cal N} \gamma$ 
for ${\cal N}= ^{12}\!C$ and $E_\nu=1\,{\rm GeV}$. 
}
\end{center}
\end{figure}

The different components of the coherent cross section
are depicted in Fig.~{\ref{fig:coherent}}.  
Distributions in photon energy and photon angle for $1\,{\rm GeV}$ incident
neutrino energy are displayed in Fig.~\ref{fig:distrib}, for each of 
the Compton-like, $\Delta$-induced and $\omega$-induced neutrino cross sections.   
The upper limit of the band in Fig.~\ref{fig:coherent} is obtained by 
combining the effects of $\omega$ and $\Delta$ into an effective coupling $g_\omega^{\rm eff}$, 
and ignoring the resonant factor (\ref{resenhance}).  

For $\Delta$, inclusion of the
subleading vector component of the weak current leads to a correction 
of the matrix element (\ref{omegam2}): 
\begin{multline}\label{gammasub}
\bar{\Sigma}|{\cal M}|^2 \propto e^2(1-e) \bigg\{ 
1-xz \\
\pm {E\over m_N} {(1+a_p-a_n)(1-2s_W^2)\over 2g_A}
\left[ (2-e)(1-y) + (x-z)(x-(1-e)z) \right] + \dots \bigg\} \,,
\end{multline}
where the $\pm$ refer to neutrino and antineutrino cross sections.  
Note that although $E$ is of order $1\,{\rm GeV}$, the small-recoil
condition enforced by the heavy nucleus ensures that the correction 
terms involve small momentum transfers.  This leads to a suppression 
of the correction terms in both the soft (\ref{soft}) and 
collinear (\ref{collinear}) limits.   

The estimates presented here differ from previous calculations of coherent
photon production in neutrino-nucleus scattering, which were focused
at large energy.   
In \cite{Gershtein:1980wu}, $A^2$ scaling was assumed at
large energy, which is incorrect; the formulas presented there
for the energy and angular distributions also do not have the correct
low energy limit.  
In \cite{Rein:1981ys}, the correct low-energy limit is not 
reproduced for contributions arising from 
the vector component of the weak neutral current.   
For example, the cross section does not reproduce 
the infrared singularity that must be present to describe 
bremsstrahlung emission.  
At higher energy, the strategy 
of relating the weak vector component of the cross section to
the forward Compton amplitude is also problematic.  
The appearance of different isospin combinations for 
photon versus $Z^0$ couplings 
invalidates the assumption of a universal ratio of neutrino and photon cross 
sections mediated by different resonances.   
At low energy, this
assumption is avoided in the present work by explicit consideration 
of the dominant ($N$, $\Delta$) intermediate states. 
The weak axial-vector cross section calculated in \cite{Rein:1981ys}
corresponds roughly to the effects of our $t$-channel $\omega$
exchange.   However, the analogy used there between an effective
$Z^* \gamma \gamma^*$ vertex~\cite{Rosenberg:1962pp} 
and a $Z^* \gamma \omega^*$ vertex 
is incorrect.
In the former case ($\gamma^*$), the amplitude must vanish when 
the $p^2$ of the photon goes to zero, due to gauge invariance and 
electromagnetic current conservation%
\footnote{
Equivalently, by the Landau-Yang theorem forbidding the decay of
a massive vector into two identical massless vectors.
}.   
In the latter case ($\omega^*$), the amplitude does {\it not} vanish
as the $p^2$ carried by $\omega$ goes to zero, due to {\it nonconservation} 
of the baryon current.  
This is a reflection of the counterterm structure studied systematically 
in \cite{HHH1}.  
It turns out that (apart from a misplaced factor of $2$ in relating 
$\gamma N\to \pi^0 N$ and the $\pi^0 N\to \gamma N$ cross sections),
their final cross section corresponds to the expression obtained here
after integrating out $\omega$.   
This results from a combined argument involving  
the (incomplete) $Z^* \gamma \gamma^*$ analogy and an appeal to 
PCAC~\cite{Adler:1964yx}.  
The PCAC argument brings in further complications from
$\pi^0$ absorption in the $\gamma N \to \pi^0 N$ cross section, that are avoided by direct 
calculation.   
The axial-vector excitation of $\Delta(1232)$ is also omitted from \cite{Rein:1981ys}, 
which is not justified in the GeV energy range.  

\subsubsection{Coherent single pion production}

While it is not a focus of the present paper to explore the phenomenology 
of single pion production~\cite{pi0}, this process is closely linked to that of
single photon production.  It is instructive to mention 
some of the similarities and differences that are encountered.  

$\Delta$ should give
the most prominent contribution to coherent single-pion production 
at low energies.  Contributions to the matching condition (\ref{eq:piNc}) 
from effective scalar ($\sigma$) degrees of freedom or from 
other excited nucleon states ($N^*$) are subdominant, both from 
formal $1/N_c$ counting arguments, and from plausible numerical 
estimates~\cite{Bernard:1996gq}.  
In addition to this normalization, the $\Delta$ contribution
leads to a resonant enhancement, in analogy to (\ref{resenhance}).  

Compared to (\ref{gammasub}), the coherent single pion cross section mediated by $\Delta$ 
involves 
\begin{multline}\label{pisub}
\bar{\Sigma}|{\cal M}|^2 \propto e^2(1-e) \bigg\{ 
1-y+2xz \\
\pm {E\over m_N} {(1+a_p-a_n)(1-2s_W^2)\over g_A}
\left[ (2-e)(1-y) - (x-z)(x-(1-e)z) \right] + \dots \bigg\} \,.
\end{multline} 
In contrast to (\ref{gammasub}),  
Eq.(\ref{pisub}) shows that the leading term for coherent pion production through the $\Delta$ resonance, 
$\propto 1-y+2xz$, is not suppressed in the collinear limit.  This, together with 
the relative minus sign in the two parts of the subleading term, implies a 
smaller difference in neutrino and antineutrino cross sections for coherent 
$\pi^0$ production compared to single photon production.   
(As already noted, the collinear region is not the dominant one at energies large compared
to $(m_\Delta^2-m_N^2)/2m_N \approx 300\,{\rm MeV}$, mitigating the suppression of
photon production.)

Let us make 
a few remarks on normalizing single photon production to $\pi^0$ production.  
If the cross sections for the two processes are related, 
measurement of the relatively 
abundant production of $\pi^0$ allows a constraint to
be placed on the single photon process.  For example, this 
procedure is used by the MiniBooNE collaboration to place bounds on single photon events as
a background to $\nu_e$ appearance measurements~\cite{MiniBooNE,AguilarArevalo:2008rc}.
For incoherent scattering off individual nucleons, this procedure appears
reasonable as a first approximation, provided the effects of pion reabsorption
and rescattering can be accounted for.  
As indicated by 
Fig.~\ref{fig:err} and the accompanying discussion, 
single photon production is not very sensitive
to offshell modifications; roughly speaking, an onshell $\Delta(1232)$ is produced, and the 
relative number of single-photon and $\pi^0$ events should approximately 
follow the decay branching ratios of $\Delta\to N\pi^0$ and $\Delta\to N \gamma$. 

For the coherent interaction however, a number of complications enter into
an attempt to normalize single-photon production to the $\pi^0$ rate.   
In contrast to the incoherent process, coherence enforces nontrivial constraints
on the phase space for final-state particles, e.g. via the ansatz (\ref{ff}). 
Note that before the constraints are imposed, the total cross sections retain 
the relative normalization that would be predicted from the decay of an 
onshell $\Delta$ in the nonrelativistic limit%
\footnote{ 
The absolute normalization of the coherent cross section in the 
nonrelativistic limit corresponds to $2/3$ of 
the total cross section for 
$\nu p \to \nu \Delta^+$, 
since there is a nonzero amplitude to flip the nucleon spin.    
}.   This 
can be verified by integrating (\ref{gammasub}) and (\ref{pisub}) over 
the final-state phase space, 
$\int dx\,dy\,de\, e(1-e)$ (or $\int dx\,dz\,de\, e(1-e)$). 
Also in the soft limit (\ref{soft}) where $e\to 0$, 
the normalization is still retained in the presence 
of the coherent form factor (\ref{ff}), as can be seen by taking $y=1$, $x=z$ 
in (\ref{gammasub}) and (\ref{pisub}), and integrating $\int dx$.    
Note that the cross sections in this limit 
have very different distributions in phase space, 
with the photons distributed according to $1-x^2$, and the pions distributed according to 
$2x^2$.  

Although the soft region dominates in the asymptotic $bE^2\to\infty$ limit, 
it is far from being precisely satisfied at $E\sim 1\, {\rm  GeV}$ on relatively
small (e.g. $^{12}$C) nuclei.   
In between the limits $bE^2\to 0$ and $bE^2\to\infty$, 
the coherent single-photon and $\pi^0$ cross sections mediated by $\Delta$ 
are not simply related.    
The strong interaction of an emitted pion with the nucleus 
further complicates any simple comparison of $\pi^0$ and $\gamma$ production.  
Given these observations, with sufficient experimental data 
it may be useful to work in the opposite direction, i.e., 
to use the single-photon mechanism to isolate these effects, since the
photon events do not suffer significant final-state interactions.   

\section{Summary and discussion \label{sec:conclusion}}

This paper surveyed single-photon production in neutrino-nucleon 
scattering in the $E_\nu \sim 1\,{\rm GeV}$ energy range.  
Several mechanisms come into play for both single-nucleon and 
coherent processes. 
At low energy, the effects are
organized by a chiral lagrangian expansion, suitably extended to
include the effects of general neutral electroweak gauge fields.   
For applications above a few hundred MeV, the
effects of the dominant resonances in
each channel were incorporated.  

The results should provide a useful guide to
the experimentalist dealing with single-photon signals and backgrounds.
Several features of the analysis are also intriguing from a theoretical perspective. 
These include the notion of color-enhancement for the $\Delta$-induced
interaction; novel coherent-resonant phenomena reminiscent of the super-radiant effect
in atomic physics; and the connection of the $\omega$-induced interaction to
the baryon current anomaly as noted in Ref.~\cite{HHH2}. 

Single-photon events are an important background in experimental configurations
such as MiniBooNE and T2K
searching for $\nu_e$ appearance in a $\nu_\mu$ beam.   This happens because 
a $\nu_\mu$ induced photon shower can be mistaken for the $\nu_e$ induced 
electron signal.   Let us comment briefly on strategies to constrain these
backgrounds. 
The $\Delta$ resonance plays a leading role.  We noticed
that the effects of uncertain offshell parameters are formally suppressed
but have significant impact at low-energy.  However, above several hundred MeV the 
rate of single photon production is found to be less sensitive to these offshell 
modifications.   Thus, e.g., the procedure employed by MiniBooNE to normalize 
(incoherent) $\Delta\to N\gamma$ events in terms of the measured $\Delta \to N\pi^0$ rate 
appears reasonably justified as a first approximation, provided that $\pi^0$
absorption and rescattering effects can be reliably accounted for%
\footnote{
For related discussion on the issue of final-state pion interactions, 
see \cite{AguilarArevalo:2008rc,Leitner:2008fg}. 
}.   
It would however be straightforward to calculate this rate directly, avoiding
complications of the strongly-interacting pion inside the nucleus.   
Effects of Compton scattering and $t$-channel $\omega$ exchange could likewise
be simulated directly for the incoherent process. 

The $\Delta$ resonance also plays a prominent role in the coherent process.
Comparing to the analogous $\pi^0$ production is more difficult for this case.   Besides the important
effects of pion-nucleus interactions, the small-$Q^2$ constraints imposed by
nuclear form factors blurs any simple relation between $\pi^0$ and $\gamma$ 
production.  In particular, the $\gamma$ production will have a broader (less forward) 
angular distribution, and a larger vector-axial interference leading to a larger
difference between $\nu$ and $\bar{\nu}$ cross sections.    
It appears that the $\omega$-induced single-photon production plays a subdominant role 
at $\sim 1\,{\rm GeV}$, being suppressed by form factors and recoil compared
to naive estimates~\cite{HHH2}.   Interference effects between $\omega$ and $\Delta$ contributions
will be significant at low energy where the respective amplitudes 
are described by the same effective operator.  
Note also that although the coherent $\Delta$ 
contribution saturates above $\sim 1\,{\rm GeV}$,  in our vector dominance model 
the $\omega$ contribution continues to grow as $E^2$ up to a parametrically 
larger scale, $E\sim A^{1/3} m_{\rm axial}^2 / (1\,{\rm GeV})$; it may thus have
a significant role to play at intermediate (few GeV) energies.   
Similarly, the correction to coherent $\pi^0$ production in 
(\ref{pisub}) from $\omega$ becomes more significant at large energy.
The Compton-like contribution to the coherent single-photon production 
was found to be small compared to the other mechanisms.  
It is interesting to note however that this process would provide an indirect means of observing 
coherent neutrino-nucleus scattering~\cite{Freedman:1973yd}, via initial and
final-state radiation.    

It is interesting to consider whether new mechanisms 
of single photon production, in particular coherent processes, 
could explain an excess of events observed by MiniBooNE~\cite{AguilarArevalo:2008rc}.  
From Figs.~\ref{fig:coherent} and \ref{fig:distrib}, the $\Delta$-induced component
appears to be the largest effect; it is interesting that this component also has 
photon energy and angular distributions
most closely resembling the excess.  
From the simple estimates presented here, the 
per-nucleon cross section for the coherent mechanism is similar in 
size to the incoherent case, and is not directly constrained by the analog 
$\pi^0$ production rate.  
In addition to this coherent-resonant effect, 
the $\omega$-mediated process appears to be subdominant but non-negligible, 
and will add constructively to the $\Delta$-mediated amplitude at low energy. 
Combined with the incoherent processes, and within large uncertainties, there appears
to be a sufficient number of photons to cover the excess.    
More definitive statements would require further study of nuclear effects
and detector efficiencies.  
It would be helpful to gather sufficient statistics to obtain 
cross sections for $\bar{\nu} N \to \bar{\nu} N \gamma$~\cite{AguilarArevalo:2009xn}. 
Of course, distinguishing photon and
electron events at the detector level would provide a decisive 
discrimination between different backgrounds.  
Similar cross sections should be measured by the T2K experiment. 
At lower energy, it is interesting to consider whether the analysis 
of signal events at the LSND experiment~\cite{Athanassopoulos:1996jb,Athanassopoulos:1997pv} 
could have been influenced by a feed-down from decay-in-flight 
neutrinos that produce single photons.  

It is interesting to pursue a more systematic classification in 
terms of invariant form factors for the general two-current 
electroweak matrix element.  This would allow the definition of 
physical observables incorporating more information than 
the total cross sections.  
Such classification would generalize known results for
the case of two vector currents 
(i.e. two photons)~\cite{Bardeen:1969aw,Perrottet:1973qw,Tarrach:1975tu} to the case 
where one or both of the currents is axial-vector. 
In addition to enforcing constraints of discrete symmetries (parity and time 
reversal) and gauge invariance, 
a useful classification must account for the infrared 
singularities encountered in certain subprocesses.  
However, it is not obvious that such a classification would provide 
immediate physical insight, since an underlying dynamical 
model or small-parameter expansion is needed to parameterize both
the normalization and shape of the invariant amplitudes%
\footnote{
For the case of real photon production from either the vector 
or axial-vector weak current, 12 invariant amplitudes correspond
to the independent helicity amplitudes of $N Z^*\to N\gamma$ 
when the leptonic current is conserved, 16 when lepton masses are 
considered. 
}.
The present paper has instead focused on the systematic expansion of the chiral
lagrangian at low energies, and used a phenomenological ansatz to perform
a modest extrapolation into the GeV energy range.  
It would be interesting to investigate the invariant amplitudes in more
detail and to look more systematically at higher energies. 
These aspects of formalism are beyond the scope of the present paper, 
but could be relevant to experiments with higher-energy neutrino beams. 
It is also interesting to consider the relevant interactions in the
context of string-inspired models of QCD~\cite{Domokos}.  

The relation between the baryon current anomaly and measurements of coherent
photon interactions is more intricate than envisaged in Ref.~\cite{HHH2}, 
which motivated the present work.  $\Delta$ and $\omega$ 
match onto the same effective operator at 
energies $E \ll m_\omega, m_\Delta-m_N$.   
The amplitudes constructively interfere, with $\Delta$ appearing
to give a larger contribution.   
This clouds 
the intriguing connection between low-energy electroweak probes 
and the standard model baryon current anomaly.   
However, at the practical level, the
effective coupling $g^{\rm eff}_\omega$ 
should be larger than considered in \cite{HHH2}.   
A more in-depth study of possible astrophysical applications is left to future 
work.  

\vskip 0.2in
\noindent
{\bf Acknowledgements}
\vskip 0.1in
\noindent
I thank J.~Harvey and C.~Hill for useful discussions and collaboration in the
early stages of this work.  
This research was supported in part 
by the U.S.~Department of Energy grant DE-AC02-76CHO3000. 

%%%%%%%%%%%%%%%%%%%%%%%%%%%%%%%%%%%%%%%%%%%%%%%%%%%%%%%%%%%%%%%%%%%%%%%
%\newpage

\begin{appendix}
\section{Compton amplitude to subleading order \label{sec:app}}

This appendix presents the result of expanding 
the diagrams of Fig.~\ref{fig:compton} to subleading order in the $1/m_N$ expansion.  
This allows a direct comparison of the effects of the pCS operators to 
those of Compton scattering, as discussed after (\ref{nrlimit}).  
Define: 
\be
\label{eq:poldef}
{i \cal M} = { - {ieg_2 \over 2c_W} }\epsilon^{(\gamma)*}_\mu \epsilon^{(Z)}_\nu T^{\mu\nu} \,.
\ee
Working in units where $E_\gamma =q^0 =1$, the result is:
\begin{align}\label{subleadingcompton}
T^{00} &= \frac{1}{m_N}\bigg[ 
F_1 C_V ( 2\bm{q}\cdot \bm{p} ) 
+ F_1 C_A (-2\bm{\sigma}\cdot \bm{q} ) \bigg] \nl
& + \frac{1}{m_N^2} \bigg[ 
F_1 C_V \left( 2\bm{q}\cdot \bm{p} \bm{q} \cdot (\bm{k}+\bm{k'}) 
- i \bm{\sigma}\cdot \bm{q}\times \bm{p} \right) 
+ F_1 C_2 \left( -i \bm{\sigma}\cdot \bm{q}\times \bm{p} \right) \nl
& \quad 
+ F_1 C_A \left( -\bm{q}\cdot \bm{p} \bm{\sigma}\cdot (\bm{k}+\bm{k'}) 
- \bm{q}\cdot (\bm{k}+\bm{k'}) \bm{\sigma} \cdot \bm{q} 
\right) 
+ F_2 C_V \left( - i\bm{\sigma} \cdot \bm{q}\times \bm{p} \right) \bigg] 
\,,\nl
T^{i0} &= 
 \frac{1}{m_N} \bigg[ 
F_1 C_V ( 2p^i ) + F_1 C_A (-2\sigma^i ) \bigg] \nl
&
+ \frac{1}{m_N^2} \bigg[ 
F_1 C_V \left( \bm{q}\cdot (\bm{k}+\bm{k'}) p^i 
+ \bm{q}\cdot \bm{p} (k^i + k^{'i}) 
- i (\bm{p} \times \bm{\sigma})^i 
+ i \bm{q}\cdot \bm{p} (\bm{q} \times \bm{\sigma})^i 
\right) \nl
&\quad
+ F_1 C_2 \left( - i(\bm{p} \times \bm{\sigma} )^i \right) 
+ F_1 C_A \left( -\bm{q}\cdot (\bm{k}+\bm{k'}) \sigma^i 
- p^i \bm{\sigma} \cdot (\bm{k}+\bm{k'}) \right) \nl
&\quad 
+ F_2 C_V \left( -i (\bm{p}\times \bm{\sigma} )^i 
+ i \bm{q}\cdot \bm{p} (\bm{q}\times  \bm{\sigma})^i \right) 
+ F_2 C_A \left( (k^i + k^{'i}) \bm{q}\cdot \bm{\sigma} 
- \bm{q} \cdot (\bm{k}+\bm{k'}) \sigma^i \right) \bigg] \,,
 \nl
T^{0j} &=  \frac{1}{m_N}\bigg[ 
F_1 C_V (2 q^j ) + F_1 C_A (-2\bm{q}\cdot \bm{p} \sigma^j ) \bigg] \nl
& + \frac{1}{m_N^2} \bigg[ 
F_1 C_V \left( 
\bm{q}\cdot \bm{p} (k^j + k^{'j}) 
+ \bm{q}\cdot (\bm{k}+\bm{k'}) q^j 
+ i(\bm{q}\times \bm{\sigma})^j 
- i\bm{p}\cdot \bm{q} (\bm{p}\times \bm{\sigma})^j \right) \nl
&\quad 
+ F_1 C_2 \left( -i\bm{q}\cdot\bm{p} (\bm{p}\times\bm{\sigma})^j 
+ i(\bm{q}\times \bm{\sigma})^j \right) \nl
&\quad
+ F_1 C_A \left( -q^j \bm{\sigma}\cdot (\bm{k}+\bm{k'}) 
+ (1-2\bm{q}\cdot\bm{p}) \bm{q}\cdot (\bm{k}+\bm{k'}) \sigma^j \right) \nl
&\quad
+ F_2 C_V \left( i (\bm{q} \times \bm{\sigma} )^j \right) 
+ F_2 C_A \left( i(\bm{q}\times \bm{p} )^j 
+ (k^j + k^{'j}) \bm{q}\cdot \bm{\sigma} 
- q^j (\bm{k}+\bm{k'})\cdot\bm{\sigma} \right)
\bigg]  \,,\nl
T^{ij} &= \frac{1}{m_N}\bigg[
F_1 C_V ( 2\delta^{ij} ) 
+ F_1 C_A\left( -2p^i \sigma^j + 2 ( \delta^{ij} \bm{q}\cdot \bm{\sigma} 
- q^j \sigma^i ) \right)
+ F_2 C_A \left( 2( \delta^{ij}\bm{q}\cdot \bm{\sigma} - q^j \sigma^i) \right)
\bigg] \nl
& + \frac{1}{m_N^2} \bigg[
F_1 C_V \big( i\epsilon^{ijr} \sigma^r (-1+\bm{p}\cdot\bm{q} - \bm{k}\cdot\bm{k'}) 
+ k^{'i} k^{'j} - k^i k^j + k^{'i} q^j + q^i k^{'j} + q^i k^j + k^i q^j \nl
&\quad + i\epsilon^{jrs}\sigma^s ( -(k+k')^r(k+k')^i -p^r p^i + q^r q^i )
+ i\epsilon^{irs} \sigma^s ( (k+k')^r (k+k')^j - p^r p^j + q^r q^j ) \big) \nl
&\quad 
+ F_1 C_2 \big( -i\epsilon^{ijr}\sigma^r 
+ i\epsilon^{jrs}p^r( \delta^{is} \bm{\sigma}\cdot \bm{q} - p^i \sigma^s 
- q^s \sigma^i ) \big) \nl
&\quad
+ F_1 C_A \big( -(1-\bm{p}\cdot\bm{q})i\epsilon^{ijr}q^r 
-\delta^{ij} \bm{\sigma}\cdot(\bm{k}+\bm{k'}) 
+ \bm{q}\cdot(\bm{k}+\bm{k'}) (q^j \sigma^i - \delta^{ij}\bm{q}\cdot\bm{\sigma})
\nl
&\qquad\qquad
+ \sigma^j( -\bm{q}\cdot(\bm{k}+\bm{k'}) q^i 
-2\bm{q}\cdot\bm{k'} k^{'i} + 2\bm{q}\cdot\bm{k} k^i ) \big) \nl
&\quad
+ F_2 C_V \big( -i\epsilon^{ijr}\sigma^r + i\epsilon^{irs} q^r 
( -\delta^{sj}\bm{\sigma}\cdot\bm{p} + q^j\sigma^s + p^s \sigma^j ) \big) 
\nl
&\quad
+F_2 C_2 \big( 2i \epsilon^{jrs} p^r ( \delta^{is} \bm{\sigma}\cdot\bm{q} 
- q^s \sigma^i ) \big) \nl
&\quad
+F_2 C_A \big( (\bm{q}\cdot\bm{p} q^r - p^r )i\epsilon^{ijr} 
+ (k+k')^j \sigma^i - \delta^{ij} (\bm{k}+\bm{k'})\cdot\bm{\sigma} \big) 
\bigg]  \,.
\end{align} 
The notation $C_2=F_V^{2\,, {\rm weak}}(0)$ is used.  It is straightforward to check
that $T^{00} = q^i T^{i0}$ and $T^{0j} = q^i T^{ij}$ as required by gauge invariance.  
For physical polarization states of the photon, we should
take $\mu=i$ in (\ref{eq:poldef}).  
Interactions involving $C_A$ are 
then seen to be either spin-dependent, or vanish when $\bm{p}\to\bm{q}$; 
that is, they affect either the spin or the momentum of the struck nucleon. 

\end{appendix}

\end{document}